\documentclass[aps,pre,reprint,floatfix,superscriptaddress]{revtex4-1}
\pdfoutput=1
\usepackage{amsmath}
\usepackage{amssymb}
\usepackage{graphicx}
\usepackage[colorlinks=true,allcolors=blue]{hyperref}

\newcommand{\sym}[1]{\hat{#1}}
\newcommand{\asym}[1]{\check{#1}}
\newcommand{\species}[1]{#1_s}
\newcommand{\rhopol}{\rho_\text{pol}}
\newcommand{\cS}{c_\text{S}}
\newcommand{\vA}{v_\text{A}}
\newcommand{\alfven}{Alfv\'{e}n}
\newcommand{\wS}{\tilde{\omega}_\text{S}}
\newcommand{\wG}{\tilde{\omega}_\text{G}}
\newcommand{\Rm}{R_\text{m}}
\newcommand{\gammares}{\gamma_\text{res}}
\newcommand{\wGAM}{\omega_\text{GAM}}
\newcommand{\wTAE}{\omega_\text{TAE}}
\newcommand{\wBAAE}{\omega_\text{BAAE}}
\newcommand{\vbf}[1]{\boldsymbol{\mathrm{#1}}}  
\newcommand{\kgeo}{\mathcal{K}}
\newcommand{\kpar}{k_\parallel}
\newcommand{\btilde}{\tilde{\beta}}
\newcommand{\wtilde}{\tilde{\omega}}
\newcommand{\dtilde}{\tilde{\delta}}
\newcommand{\etilde}{\tilde{\varepsilon}}
\newcommand{\stilde}{\tilde{s}}

\newcommand{\rhotilde}{\tilde{\rho}}
\newcommand{\Psib}{\Psi_\text{b}}
\newcommand{\qcyl}{\tilde{q}}
\newcommand{\qnot}{\tilde{q}_\ddagger}
\newcommand{\xiA}{\xi^\text{A}}
\newcommand{\xiS}{\xi^\text{S}}
\newcommand{\xiAtilde}{\tilde{\xi}^\text{A}}
\newcommand{\xiStilde}{\tilde{\xi}^\text{S}}
\newcommand{\Diag}[1]{D_{#1}}
\newcommand{\Liagp}[1]{w^{+}_{#1}}
\newcommand{\Liagm}[1]{w^{-}_{#1}}
\newcommand{\Uiagp}[1]{u^{+}_{#1}}
\newcommand{\Uiagm}[1]{u^{-}_{#1}}

\begin{document}

\title{
High-order geodesic coupling of shear-\alfven{} and acoustic continua in
tokamaks}

\date{\today}

\author{Paulo Rodrigues}
\affiliation{Instituto de Plasmas e Fus\~{a}o Nuclear, Instituto Superior
T\'{e}cnico, Universidade de Lisboa, 1049-001 Lisboa, Portugal.}

\author{Francesca Cella}
\affiliation{Instituto de Plasmas e Fus\~{a}o Nuclear, Instituto Superior
T\'{e}cnico, Universidade de Lisboa, 1049-001 Lisboa, Portugal.}
\affiliation{Dipartimento di Energia, Politecnico di Milano, Via Ponzio 34/3,
20133 Milan, Italy.}

\begin{abstract}
High-order plasma shaping (mainly elongation and shift, as opposed to low-order
toroidicity) is shown, under certain conditions, to open gaps in the coupled
shear-\alfven{} and acoustic continua at frequencies significantly above the
values predicted by previous theories. Global eigenmodes in these gaps, which
lie between those of geodesic acoustic modes (GAMs) and toroidicity-induced
\alfven{} eigenmodes (TAEs), are found unstable to hot-ion populations typical
of tokamak operation, whilst their fundamental resonances with circulating
particles are shown to take place at velocities near the geometric mean of the
\alfven{} and sound speeds. Therefore, such eigenmodes are expected to be
observed near the predicted frequencies at operating tokamaks, playing a still
unexplored role in magnetohydrodynamic spectroscopy as well as in the stability
of next-step fusion experiments.
\end{abstract}

\maketitle

\section{Introduction}

Continuous spectra of the magnetohydrodynamics (MHD) operator are central to a
variety of phenomena dominated by inhomogeneous magnetic
fields~\cite{uberoi.1972, grad.1973, goedbloed.1975}, from astrophysical plasmas
to fusion devices. Their origin lies on vanishing coefficients in the eigenvalue
equation
\begin{equation}
  \mathcal{F}(\vbf{\xi}) + \mu_0 \rho \omega^2 \vbf{\xi} = 0
\label{eq:eigenvalue.problem}
\end{equation}
for small plasma displacements $\vbf{\xi} e^{-i \omega t}$, where $\mu_0$ is the
magnetic constant and $\rho$ is the mass density, with
\begin{multline}
  \mathcal{F}\bigl( \vbf{\xi} \bigr) =
      \bigl( \nabla \times \vbf{B} \bigr) \times
      \bigl[ \nabla \times \bigl( \vbf{\xi} \times
      \vbf{B} \bigr) \bigr] \\
    + \bigl[ \nabla \times \nabla \times
      \bigl( \vbf{\xi} \times \vbf{B} \bigr) \bigr] \times \vbf{B} \\
    + \mu_0 \nabla \bigl( \gamma P \, \nabla \cdot \vbf{\xi} +
      \vbf{\xi} \cdot \nabla P \bigr)
\label{eq:force.operator}
\end{multline}
the ideal-MHD operator~\cite{bernstein.1958, goedbloed.1975}, $P$ and $\vbf{B}$
the equilibrium pressure and magnetic field, while $\gamma = 5/3$. As a simple
but rather conveying example, let $\rho$ be uniform, $\nabla \cdot \vbf{\xi} =
0$, and $\vbf{B} = B(x) \vbf{e}_z$, which turns
Eq.~\eqref{eq:eigenvalue.problem} into~\cite{uberoi.1972}
\begin{equation}
  \frac{d}{dx}\biggl[
    \Bigl( \omega^2 - k_\parallel^2 \vA^2 \Bigr)
      \frac{d}{dx}\tilde{\xi}_x \biggr]
    - k^2 \Bigl( \omega^2 - k_\parallel^2 \vA^2 \Bigr) \tilde{\xi}_x = 0,
\label{eq:inhomogeneous.example}
\end{equation}
where $\vbf{\xi} = \tilde{\vbf{\xi}}(x) \exp[i(k_y y + k_z z)]$, $\kpar =
\vbf{k} \cdot \vbf{b}$ is the parallel wave number, $\vbf{b} = \vbf{B}/B$ is the
field versor, and $\vA^2 = B^2 / (\mu_0 \rho)$ is the squared \alfven{} speed.
Near any $x'$, the eigenvalue $\omega^2(x') = k_\parallel^2(x') \vA^2(x')$
defines a singular solution $\tilde{\xi}_x \propto K_0( k | x - x' |)$, with
$K_0(x) = - \ln x + \cdots$ the modified Bessel function of the second
kind~\cite{abramowitz.1972}. Unlike the discrete spectra of Sturm-Liouville
operators (whose coefficients do not vanish in their domain),
Eq.~\eqref{eq:inhomogeneous.example} produces a set of eigenvalues $\omega(x)$
that depend continuously on the variable $x$ along the inhomogenity direction
(i.e., a continuum). Likewise, eigenvalues in the continuum of the more general
Eq.~\eqref{eq:eigenvalue.problem} define singular waves that travel along field
lines at $\vA$ (transverse $\vbf{b}\times\vbf{\xi}\times\vbf{b}$ or
shear-\alfven{} waves) or at the sound speed (parallel $\vbf{\xi}\cdot\vbf{b}$
or acoustic waves), the square of the latter being $\cS^2 = \gamma
P/\rho$~\cite{uberoi.1972, grad.1973, goedbloed.1975}. Regardless of their
polarisation, singular continuum waves are known to be strongly
damped~\cite{tataronis.1973, grossmann.1973}.

On tokamaks, $k_\parallel = 0$ at rational surfaces while $\vA$ grows unbounded
as $\rho \rightarrow 0$ at the edge, and continuum frequencies should thus span
the range $0 \leqslant \omega < \infty$~\cite{grad.1973, appert.1974}. However,
the field $\vbf{B}(\Psi, \vartheta)$ depends on a poloidal angle $\vartheta$
(besides $2\pi \Psi$, the poloidal-field flux labelling magnetic surfaces) and
the consequent periodicity of the refractive index opens frequency gaps in the
continua (i.e., forbidden bands)~\cite{rayleigh.1887, cheng.1986, zhang.2008}
where traveling singular waves are replaced by non-singular discrete \alfven{}
eigenmodes (AEs). Streaming along field lines at speeds close to $\vA$,
fusion-born $\alpha$-particles or other energetic ions produced by the heating
systems may resonantly lose energy to these potentially less damped
AEs~\cite{rosenbluth.1975, fu.1989, betti.1992}, leaving the plasma core in the
process. Such AE-induced transport of very hot near-\alfven{}ic particles (i.e.,
with $v \sim \vA$) may hinder the operation of future fusion reactors (burn
quench, wall damage, etc.)~\cite{fasoli.2007} and, being so, research on ion-AE
interactions~\cite{heidbrink.1994, gorelenkov.2014, lauber.2013} have focused
mostly on gaps in the shear-\alfven{} (SA) continuum, all near or above the
frequency $\wTAE = \vA/(2qR_0)$ of toroidicity-induced AEs (TAEs), with $q \sim
1$ the safety factor and $R_0$ the torus major radius~\cite{cheng.1986,
betti.1991, heidbrink.2008}.

Experimental evidences of unstable AEs with frequencies $\omega \lesssim \wTAE$
have raised the interest for gaps in the acoustic continuum also, particularly
when the plasma beta $\beta \sim \cS^2/\vA^2$ lies in the range $2\% \lesssim
\beta \lesssim 4\%$~\cite{heidbrink.1993, turnbull.1993, heidbrink.1999,
huysmans.1995, heidbrink.2021}. Frequency gaps below $\wTAE$ were found in
numerically computed continua, at first using the slow-sound limit $\omega \gg
\cS/R_0$~\cite{chu.1992, turnbull.1993} and then the full set of linear MHD
equations~\cite{huysmans.1995} to describe the SA-acoustic coupling, their width
being proportional to $\beta$ in either case.  Inside such gaps, $\beta$-induced
AEs (BAEs) were also computed~\cite{turnbull.1993, huysmans.1995}, their squared
frequency scaling with $\cS^2$ and thus with the plasma
temperature~\cite{huysmans.1995}. Further numerical
simulations~\cite{cheng.2019, kramer.2020} found that the coupling between SA
and acoustic continua, as well as the corresponding frequency gaps and AEs
(therein termed \alfven{}-Slow eigenmodes or ASEs) is a robust and ubiquitous
feature of tokamak plasmas, being present for a large variety of $\beta$ values,
$q$ profiles (monotonic, reversed, or weakly sheared), and equilibrium shaping.

Unlike gaps in the SA continuum, for which analytical insight about the two-wave
coupling and the resulting AE location and frequency is readily
available~\cite{cheng.1986}, acoustic-wave couplings are far more complex.
Initial analytical estimates for gap frequencies and loci were limited to
cylindrical equilibria and decoupled continua~\cite{huysmans.1995}, being thus
of modest accuracy and practical utility. Keeping circular magnetic surfaces but
allowing finite toroidicity was later shown to couple SA and acoustic waves near
a rational surface~\cite{holst.2000b, gorelenkov.2007, gorelenkov.2007b},
opening a gap below the sound frequency $\cS / R_0$ and lifting the SA continuum
bottom from zero up to the value
\begin{equation}
\wGAM = \bigl(\cS/R_0\bigr) \sqrt{2 + 1/q^2},
\label{eq:gam.frequency}
\end{equation}
i.e., the typical frequency of geodesic acoustic modes~\cite{winsor.1968}.
Additional gaps, all below the sound frequency, were found recently by further
analytical developments with the same circular equilibrium
model~\cite{cheng.2019}. In summary, despite long-standing numerical
evidences~\cite{huysmans.1995, cheng.2019, kramer.2020}, no gaps have thus far
been predicted analytically between the sound frequency and $\wTAE$ other than
the bottom of the uplifted SA continuum. All previous
theories~\cite{holst.2000b, gorelenkov.2007, gorelenkov.2007b, cheng.2019}
predict gap frequencies below the former, which is much lower than the latter
because $\cS^2/\vA^2 \ll 1$. AEs at such low frequencies are expected to be
strongly damped by resonant thermal ions and thus less likely to be driven
unstable. Also, the aforementioned coupling models are unable to fully explain
measurements of \alfven{}ic activity (frequency and radial position) recently
reported to have been observed in JET experiments, with frequencies lying
precisely between $\wGAM$ and $\wTAE$~\cite{rodrigues.2021fec}.

In this work, high-order plasma shaping (elongation and shift) is shown to
couple SA and acoustic continua, opening frequency gaps in the range $\wGAM <
\omega \lesssim \wTAE$. The predicted gaps lie well above the sound frequency
and are a generalisation to shaped equilibria of previous analytical
results~\cite{holst.2000b, gorelenkov.2007, gorelenkov.2007b, cheng.2019}, all
of which were obtained in the low-order limit of circular magnetic surfaces and
finite toroidicity. These earlier results are briefly reviewed in
Sec.~\ref{sec:continua.coupling} in order to grasp the need for more accurate
equilibria. An equilibrium model with plasma shaping~\cite{rodrigues.2018} is
introduced and then employed to understand how each of its shaping harmonics
contributes to couple SA and acoustic waves, eventually showing that shift and
plasma elongation alone play a significant role in tokamaks. Insight into the
coupling mechanism is developed in Sec.~\ref{sec:frequency.gaps}, where the
coupled MHD equations are expanded in powers of two small parameters: the
inverse aspect ratio and the size of the shaping harmonics. This approach allows
tractable equations for the coupled continua to be solved and produces
analytical estimates of gap frequencies and loci, as well as an existence
condition that depends on local equilibrium geometry and $q$ values. Such
analytical estimates provide useful tools to interpret eventual experimental
observations, as well as to develop MHD-spectroscopy
techniques~\cite{goedbloed.1993, fasoli.2002}. In
Sec.~\ref{sec:resonant.interactions}, global AEs in these gaps are found
unstable to hot ions below $1$~MeV for typical tokamak parameters. Fundamental
resonances with circulating ions are shown to take place near $v^2 \sim \cS
\vA$. Hence, these AEs avoid strong thermal-ion damping while tapping enough
energy from hot ions or fusion products. Their still unexplored role in the
stability of next-step fusion experiments like ITER~\cite{aymar.2002} is briefly
discussed in Sec.~\ref{sec:discussion}.

\section{
\label{sec:continua.coupling}
Continua coupling: equilibria and geodesic curvature}

As detailed elsewhere~\cite{hameiri.1981, hameiri.1985, cheng.1986},
Eq.~\eqref{eq:eigenvalue.problem} is more conveniently written as a matrix
problem whose singular continuum is the set of non-trivial solutions of the
system
\begin{equation}
\begin{bmatrix}
\frac{\omega^2}{\vA^2} +
  \frac{B}{g} \nabla_\parallel
    \bigl(\frac{g}{B} \nabla_\parallel\bigr) &
    \frac{\cS^2}{\vA^2} \frac{B^2}{g} \kgeo \\
  \kgeo & \!\!\!\!\!\!\!\!\!\! 1 +
    \frac{\cS^2}{\vA^2} + \frac{\cS^2}{\omega^2} B \, \nabla_\parallel
\bigl( \frac{1}{B} \nabla_\parallel\bigr)
\end{bmatrix}
\begin{bmatrix}
\xiA \\
\xiS
\end{bmatrix} = 0,
\label{eq:matrix.E}
\end{equation}
coupling the shear-\alfven{} $\xiA = \vbf{\xi} \cdot \vbf{B} \times \nabla\Psi /
g$ and acoustic $\xiS = \nabla \cdot \vbf{\xi}$ components of the plasma
displacement $\vbf{\xi}$, with $g = \big|\nabla\Psi\big|^2$, $\nabla_\parallel =
\vbf{b} \cdot \nabla$, while $\kgeo = 2 \vbf{\kappa} \cdot \vbf{b} \times
\nabla\Psi/B$ and $\vbf{\kappa} = \nabla_\parallel \vbf{b}$ are, respectively,
the geodesic and field-line curvatures.

In the cylindrical-equilibrium limit, $\Psi$ and $B$ do not depend on the angle
$\vartheta$, $\vbf{\kappa}$ is parallel to $\nabla \Psi$, and the field lines
are geodesics, whence $\kgeo \rightarrow 0$, $\nabla_\parallel \rightarrow i
k_\parallel$, and two decoupled continua arise from Eq.~\eqref{eq:matrix.E}
as~\cite{grad.1973, appert.1974, goedbloed.1975}
\begin{equation}
\omega^2 = k_\parallel^2 \vA^2
  \quad \text{and} \quad
    \omega^2 = k_\parallel^2 \cS^2 \big/ (1 + \cS^2/\vA^2).
\label{eq:cylindrical.continua}
\end{equation}
In general, however, $B$ and $\Psi$ depend on $\vartheta$, the harmonics in
$\xiA = e^{i n \phi} \sum_m \xiA_m(\Psi) e^{i m \vartheta}$ (and similarly for
$\xiS$, with $\phi$ the toroidal angle around the torus) become coupled, and
Eq.~\eqref{eq:matrix.E} turns into a nondiagonal algebraic system. There, each
$p$-index harmonic of the periodic $\kgeo(\Psi, \vartheta)$ couples in the same
equation $\xiA_m$ and the pair $\xiS_{m \pm p}$ for integer $p$.

Low-$\beta$, high aspect-ratio equilibria with finite toroidicity and circular
magnetic surfaces have $\kgeo \propto \sin \vartheta$ at lowest order, being
thus able to couple the three harmonics $\xiA_m$ and $\xiS_{m \pm 1}$ near a
rational surface~\cite{holst.2000b}. If the acoustic-wave term $\propto
\cS^2/(R_0^2 \omega^2)$ in Eq.~\eqref{eq:matrix.E} is dropped under the
slow-sound approximation, a SA wave $\xiA$ couples with the acoustic response
$\xiS = - \kgeo \xiA / \bigl(1 + \beta)$ to its propagation and the frequency at
the rational surface is lifted away from zero~\cite{chu.1992, turnbull.1993}.
Keeping the acoustic-wave term, and thus the two harmonics $\xiS_{m \pm 1}$,
brings the bottom of the lifted SA continuum slightly down to $\wGAM$ and opens
a gap at the beta-induced acoustic AE (BAAE) frequency~\cite{gorelenkov.2007,
gorelenkov.2007b}
\begin{equation}
\omega_\text{BAAE} = \cS/(q R_0) < \wGAM.
\label{eq:baae.frequency}
\end{equation}
The same toroidicity couples $\xiS_m$ and $\xiS_{m \pm 1}$ also, via the field
magnitude in the term $\cS^2/\vA^2$, yielding additional gaps below
$\omega_\text{BAAE}$~\cite{cheng.2019}. Further couplings are not possible
unless higher-order harmonics are considered in $\kgeo(\Psi, \vartheta)$. In the
following, plasma shaping is shown to provide such harmonics, opening additional
gaps above the frequency $\wGAM$.

\begin{figure}
\includegraphics{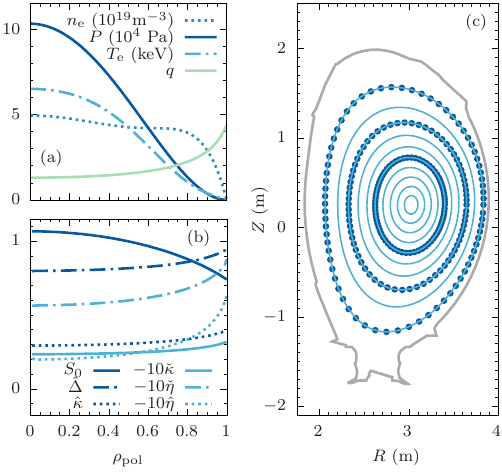}
\caption{\label{fig:jetlike.equilibrium}
JET-like equilibrium with $\varepsilon = 0.29$, $R_0 = 3$~m, $B_0 = 3.4$~T, and
$I_\text{p} = 2.3$~MA: (a) Electron density $n_\text{e}$ and temperature
$T_\text{e}$, plasma pressure $P$, and safety factor $q$; (b) fitted
coefficients $S_0$, $\sym{\Delta}$, $\sym{\kappa}$, $\asym{\kappa}$,
$\sym{\eta}$, and $\asym{\eta}$; (c) numerical (solid lines) and analytical
(large dots) magnetic surfaces.}
\end{figure}

An analytically tractable equilibrium model is built by providing a local
description of the poloidal flux~\cite{rodrigues.2018}
\begin{equation}
\Psi(r, \theta) = \Psib S_0 r^2 \Bigl[ \Theta_0(\theta) +
    \varepsilon r \Theta_1(\theta) + \varepsilon^2 r^2 \Theta_2(\theta) \Bigr]
\label{eq:localeq.flux}
\end{equation}
depending on geometric coefficients ($S_0$, $\sym{\kappa}$, $\asym{\kappa}$,
$\sym{\Delta}$, $\sym{\eta}$, $\asym{\eta}$, $\sym{\chi}$, and $\asym{\chi}$,
all constant on each magnetic surface) via
\begin{equation}
\begin{aligned}
\Theta_0(\theta) &= 1 + \sym{\kappa} \cos 2 \theta
    + \asym{\kappa} \sin 2 \theta, \\
\Theta_1(\theta) &= \sym{\Delta} \cos \theta +
    \tfrac{1}{4} \asym{\kappa} \sin \theta +
        \sym{\eta} \cos 3 \theta + \asym{\eta} \sin 3 \theta, \\
\Theta_2(\theta) &=
    \tfrac{1}{32}  \bigl( 8 \sym{\Delta} - 3 \sym{\kappa} - 3 \bigr) \\
    & \quad + \tfrac{1}{8} \bigl( 2 \sym{\eta} +
        2 \sym{\Delta} - \sym{\kappa} - 1 \bigr) \cos 2 \theta \\
    & \quad + \tfrac{1}{16} \bigl( 4 \asym{\eta} -
        \asym{\kappa} \bigr) \sin 2 \theta +
            \sym{\chi} \cos 4 \theta + \asym{\chi} \sin 4 \theta.
\end{aligned}
\label{eq:capital.thetas}
\end{equation}
Above, $r$ and $\theta$ are such that $R = R_0 \bigl( 1 + \varepsilon r \cos
\theta)$ is the distance to the torus axis, with $\varepsilon = a/R_0$, $a$ the
minor radius, and $\Psib$ the boundary flux. The field follows from $\vbf{B} =
\nabla \phi \times \nabla \Psi + B_\phi \nabla \phi$, with $B_\phi = B_0 R_0
\sqrt{1 + \varepsilon^2 S_\text{d} \Psi/\Psib}$ the covariant toroidal field,
$B_0$ the field on axis, and $S_\text{d}$ the diamagnetic coefficient. An
example is illustrated in Fig.~\ref{fig:jetlike.equilibrium} for parameters
typical of optimised scenarios at the Joint European Torus
(JET)~\cite{dumont.2018}: The equilibrium is computed
by~\texttt{HELENA}~\cite{huysmans.1991} and the local
flux~\eqref{eq:localeq.flux} is fitted to each magnetic surface to get the
geometric coefficients. These change little along $\rhopol = \sqrt{\Psi/\Psib}$
and follow the orderings
\begin{equation}
S_0, \sym{\Delta} \sim 1, \quad \sym{\kappa} \sim \varepsilon,
  \quad \text{and} \quad
    \asym{\kappa}, \sym{\eta}, \asym{\eta}, \sym{\chi}, \asym{\chi}
\lesssim \varepsilon^2.
\label{eq:geometric.ordering}
\end{equation}
Analytical magnetic surfaces are found inverting $\Psi(r, \theta)$ for a
given flux value, yielding the series~\cite{rodrigues.2018}
\begin{equation}
r(\theta) = \stilde \Biggl( \frac{1}{\Theta_0^{1/2}}
    - \frac{\Theta_1}{2 \Theta_0^2} \, \etilde
    + \frac{5 \Theta_1^2 - 4 \Theta_0 \Theta_2}{8
    \Theta_0^{7/2}} \, \etilde^2 +
 \cdots \Biggr),
\label{eq:magnetic.surface}
\end{equation}
with $\stilde^2 = S_0^{-1} \bigl( \Psi/\Psib \bigr)$ and $\etilde = \varepsilon
\stilde$.

Intricate functions of $\vbf{B}$ and $\Psi$, as is the case of $\kgeo$, are
expanded in powers of the small numbers $\etilde$ and $\dtilde \sim \etilde$,
the latter introduced here to enforce the ordering in
Eqs.~\eqref{eq:geometric.ordering} by letting $\sym{\kappa} \rightarrow \dtilde
\sym{\kappa}$, $\asym{\kappa} \rightarrow \dtilde^2 \asym{\kappa}$, and so
forth. After tracking the order of each expanded term, $\dtilde$ is replaced by
$1$ to restore physical formulae. Casting the real-valued $\kgeo$ as
\begin{equation}
\kgeo(\stilde, \theta) = \frac{\etilde}{\qcyl} \Biggl[ \kgeo_0(\stilde) +
  \sum_{p = 1}^{\infty}
    \kgeo_p^\ast(\stilde) e^{-i p \theta} + 
      \kgeo_p(\stilde) e^{i p \theta} \Biggr],
\label{eq:kgeodesic.trig}
\end{equation}
where $\qcyl = \tfrac{1}{2} S_0^{-1} a^2 B_0 / \Psib$ is the cylindrical $q$ at
lowest order, the most important coefficients $\kgeo_p$ (with $\kgeo_p^\ast$
their conjugates) are
\begin{equation}
\begin{aligned}
i \kgeo_1 &= 1 - \tfrac{3}{4} \dtilde \sym{\kappa}
  - \tfrac{1}{16} \dtilde^2 \bigl( \sym{\kappa}^2 - 12 i \asym{\kappa} \bigr)\\
  & \quad - \etilde^2 \biggl( \frac{1 + \sym{\Delta}}{\qcyl^2}
   - \frac{1 + 4 \sym{\Delta} - 6 \sym{\Delta}^2}{64}
   + \tfrac{1}{2} S_0 S_\text{d} \biggr) + \cdots,\\
i \kgeo_2 &= \tfrac{1}{4} \etilde \sym{\Delta} + \tfrac{1}{4} \etilde \dtilde
  \bigl( \sym{\Delta} - 4/\qcyl^2 \bigr) \sym{\kappa}
  + \cdots,\\
i \kgeo_3 &= - \tfrac{1}{4} \dtilde \sym{\kappa}
  - \etilde^2 \frac{1 - 2 \sym{\Delta} + 3 \sym{\Delta}^2}{32}
  + \dtilde^2 \frac{5 \sym{\kappa}^2 + 8i \asym{\kappa}}{32} + \cdots,\\
i \kgeo_5 &= \tfrac{3}{32} \dtilde^2 \sym{\kappa}^2 + \cdots,
\end{aligned}
\label{eq:curvature.coefficients}
\end{equation}
all others being ordered as $O(\etilde^3, \dtilde^3)$. A crucial step to obtain
the coefficients in Eq.~\eqref{eq:curvature.coefficients}, is the transformation
$\kgeo(r, \theta) \rightarrow \kgeo(\stilde, \theta)$ to the surface-induced
coordinate set $\{ \stilde, \theta,\phi \}$, which is achieved thanks to the
series in Eq.~\eqref{eq:magnetic.surface}.

At lowest order, with circular magnetic surfaces and toroidicity alone, one
finds $\kgeo = (2 \etilde / \qcyl) \sin \theta + \cdots$ in agreement with
earlier results~\cite{holst.2000b, gorelenkov.2007, cheng.2019}. In turn,
equilibrium shaping brings in first-order corrections to the coefficients
$\kgeo_p$ listed in Eq.~\eqref{eq:curvature.coefficients} due to finite
$\sym{\Delta}$ and $\sym{\kappa}$ (for $1 \leqslant p \leqslant 3$ only), which
are related with the Shafranov shift and plasma
elongation~\cite{rodrigues.2018}. The linear term in $\kgeo_1$ changes slightly
the already known coupling between $\xiA_m$ and the two harmonics $\xiS_{m \pm
1}$. On the other hand, those in $\kgeo_2$ and $\kgeo_3$ introduce additional
couplings with $\xiS_{m \pm 2}$ and $\xiS_{m \pm 3}$ that may open new frequency
gaps. For $p \geqslant 4$, all terms in $\kgeo_p$ are quadratic or higher powers
of $\etilde$ or $\dtilde$ and the couplings they induce are weaker, being thus
discarded.

\section{
\label{sec:frequency.gaps}
Frequency gaps: analytical estimates and numerical verification}

Near rational surfaces where the parallel wave-number vanishes, i.e.
\begin{equation}
k_\parallel R_0 = m/\qcyl + n = 0,
\label{eq:k.parallel}
\end{equation}
only the branch $\xiA_m$ of the SA continuum is close in frequency to the
acoustic branches $\xiS_{m + p}$ with $|p| = 0, 1, 2, 3$. Using the analytical
model described in Eqs.~\eqref{eq:localeq.flux} to~\eqref{eq:magnetic.surface}
to deal with equilibrium quantities and differential operators, the system in
Eq.~\eqref{eq:matrix.E} can be expanded in powers of the small parameters
$\etilde$ and $\dtilde$. Keeping only terms up to the first order, the matrix
problem is reduced to
\begin{widetext}
\begin{equation}
\begin{bmatrix}
\rhotilde \wtilde^2 - \zeta^2 &
  \kgeo_3 & \kgeo_2 & \kgeo_1 & \kgeo_0 &
    \kgeo_1^\ast & \kgeo_2^\ast & \kgeo_3^\ast \\
\kgeo_3^\ast &\Diag{-3} & \Liagp{-3} & \Uiagp{-3} & & & \\
\kgeo_2^\ast & \Liagm{-2} & \Diag{-2} & \Liagp{-2} & \Uiagp{-2} & &
  \makebox(0,0){\normalfont\Large\bfseries 0} & \\
\kgeo_1^\ast & \Uiagm{-1}& \Liagm{-1} &
  \Diag{-1} & \Liagp{-1} & \Uiagp{-1} & & \\
\kgeo_0 & & \Uiagm{0}& \Liagm{0} & \Diag{0} & \Liagp{0} & \Uiagp{0} & \\
\kgeo_1 & & & \Uiagm{1} & \Liagm{1} & \Diag{1} & \Liagp{1} & \Uiagp{1} \\
\kgeo_2 & &
  \makebox(0,0){\normalfont\Large\bfseries 0}
    & & \Uiagm{2} & \Liagm{2} & \Diag{2} & \Liagp{2} \\
\kgeo_3 & & & & & \Uiagm{3} & \Liagm{3} & \Diag{3}
\end{bmatrix}
\begin{bmatrix}
\xiAtilde_{m} \\
\xiStilde_{m - 3}\\
\xiStilde_{m - 2}\\
\xiStilde_{m - 1}\\
\xiStilde_{m}\\
\xiStilde_{m + 1}\\
\xiStilde_{m + 2}\\
\xiStilde_{m + 3}
\end{bmatrix} = 0.
\label{eq:coupling.matrix.extended}
\end{equation}
\end{widetext}
Here, mass density and frequency are normalised to their on-axis values as $\rho
= \rho_0 \rhotilde$ and $\omega =(\vA^0/R_0) \wtilde$, whereas the variables
$\xiAtilde = (\etilde/\qcyl) \xiA$ and $\xiStilde = \btilde \xiS$ are thus
defined in order to simplify the coefficients in
Eq.~\eqref{eq:coupling.matrix.extended} and to follow the ordering $\xiStilde
\sim \btilde \xiAtilde$, with $\btilde = \gamma \mu_0 P/B_0^2$. Moreover, the
principal and side diagonals are defined as
\begin{equation}
\begin{aligned}
\btilde \Diag{p} &= 1 + \btilde - \frac{1}{\rhotilde}
  \biggl(\frac{\btilde}{\wtilde^2} \biggr)
    \biggl( \zeta + \frac{p}{\qcyl} \biggr)^2, \\
\btilde w^\pm_p &= \etilde \btilde
  - \frac{\etilde}{\rhotilde} \biggl( \frac{\btilde}{\wtilde^2} \biggr)
    \biggl[ \\
  & \quad \frac{m p}{\qcyl^2}
    + \frac{3}{2} \sym{\Delta} \frac{m + p \pm 1}{\qcyl}
      \biggl( \zeta + \frac{p \pm 1/2}{\qcyl} \biggr) \\
  & \quad - \frac{1}{2 \qcyl} \biggl( \frac{1}{\qcyl} \pm \zeta \biggr)
    - \biggl( \zeta \pm \frac{1}{2 \qcyl} \biggr)
        \biggl( \zeta + \frac{m + p}{\qcyl} \biggr) \biggr], \\
\btilde u^\pm_p &= - \dtilde
  \biggl( \frac{\btilde}{\wtilde^2} \biggr) \sym{\kappa}
    \frac{m + p \pm 2}{\qcyl\rhotilde}
      \biggl( \zeta + \frac{p \pm 1}{\qcyl} \biggr),
\end{aligned}
\end{equation}
where $\zeta(\qcyl) = m/\qcyl + n$ is the dimensionless value $R_0
k_\parallel(\qcyl)$ at the radial location with safety factor $\qcyl$. Each
acoustic harmonic $\xiStilde_p$ is coupled to $\xiStilde_{p \pm 1}$ and
$\xiStilde_{p \pm 2}$ by the diagonals $w^\pm_p$ and $u^\pm_p$ that arise due to
toroidicity and shift and due to elongation, respectively. On the other hand,
all acoustic harmonics are coupled with $\xiAtilde_m$ by the geodesic-curvature
coefficients that are placed along the matrix first line and column.

The ordering of terms in Eq.~\eqref{eq:coupling.matrix.extended} is not
exclusively set by toroidal bending and plasma shaping via the small parameters
$\etilde$ and $\dtilde$. The frequency range of interest also plays a role and
different continua are found if the slow-sound approximation $\btilde /
\wtilde^2 \sim 0$ is considered~\cite{chu.1992, turnbull.1993} or if the
acoustic-frequency limit $\btilde / \wtilde^2 \sim 1$ is
taken~\cite{gorelenkov.2007, gorelenkov.2007b}. Conversely, the focus in this
work is placed on frequency values below but close to $\wTAE$, which are thus
ordered as $\btilde / \wtilde^2 \sim \etilde$ and lie between the two previous
limits. Recalling that $\xiStilde \sim \btilde \xiAtilde$, all terms arising
from the first line in Eq.~\eqref{eq:coupling.matrix.extended} become
consistently ordered if $\zeta^2 \lesssim \btilde/\etilde$ and, consequently,
one is sufficiently close to a rational surface. In turn, the side diagonals
$w^\pm_p$ and $u^\pm_p$ produce, respectively, terms ordered as $O(\etilde
\btilde, \etilde^2)$ and $O(\etilde \dtilde$) or smaller, which may be discarded
when compared with those in $\kgeo_p$ or the in main diagonal $\Diag{p}$.
Overall, for frequencies in the range $\btilde/\wtilde^2 \sim \etilde$,
Eq.~\eqref{eq:coupling.matrix.extended} simplifies to
\begin{equation}
\begin{bmatrix}
\rhotilde \wtilde^2 - \zeta^2 &
  \kgeo_3 & \kgeo_2 & \kgeo_1 & \kgeo_1^\ast & \kgeo_2^\ast & \kgeo_3^\ast \\
\kgeo_3^\ast & \Diag{-3} & & & & & \\
\kgeo_2^\ast & & \Diag{-2} & & &
  \makebox(0,0){\normalfont\Large\bfseries 0}
    & \\
\kgeo_1^\ast & & & \Diag{-1} & & & \\
\kgeo_1 & & & & \Diag{1} & & \\
\kgeo_2 & &
  \makebox(0,0){\normalfont\Large\bfseries 0}
    & & & \Diag{2} & \\
\kgeo_3 & & & & & & \Diag{3}
\end{bmatrix}
\begin{bmatrix}
\xiAtilde_{m} \\
\xiStilde_{m - 3}\\
\xiStilde_{m - 2}\\
\xiStilde_{m - 1}\\
\xiStilde_{m + 1}\\
\xiStilde_{m + 2}\\
\xiStilde_{m + 3}
\end{bmatrix} = 0.
\label{eq:coupling.matrix}
\end{equation}
Above, the line and column $p = 0$ are omitted because $\kgeo_0$ terms are
$O(\etilde \dtilde^2)$, as noticed in Eqs.~\eqref{eq:curvature.coefficients}.
Hence, $\Diag{0}$ factors out from the matrix determinant and $\xiStilde_m$
keeps a cylindrical continuum as in Eq.~\eqref{eq:cylindrical.continua}.  In
addition, one should remark that if the frequency had been ordered as $\btilde /
\wtilde^2 \sim 1$, the side diagonals would keep coupling $\xiStilde_m$ to the
other harmonics, eventually leading to frequency gaps near $\wtilde^2 \approx
\btilde / (2 \qcyl)^2$ (i.e., $\omega^2 \approx \btilde \wTAE^2$) as obtained
elsewhere~\cite{cheng.2019}.

\begin{figure}
\includegraphics[width=246pt]{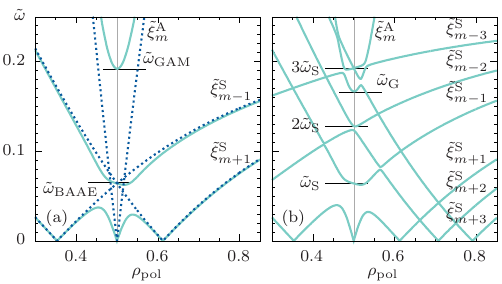}
\caption{\label{fig:simplified.continua}
Continua for $\qcyl(\rhopol) = 1 + 4 \rhopol^2$, $\rhotilde = 1$, $\btilde =
\tfrac{5}{3} \tfrac{1}{100}$, $\varepsilon = 0.3$, $S_0 = 1$, $ m = 4$, and $n =
-2$: (a) limit $\kgeo = 0$ (dots) and lowest-order coupling for circular
equilibria ($\sym{\Delta}, \sym{\kappa} = 0$, lines); (b) high-order couplings
with $\sym{\Delta} = 5$ and $\sym{\kappa} = \tfrac{1}{4}$.}
\end{figure}

Letting $\mathsf{C}$ be the coupling matrix in Eq.~\eqref{eq:coupling.matrix},
continua are found solving $\det \mathsf{C}(\rhopol, \wtilde^2) = 0$, which
factorises as
\begin{multline}
\wtilde^2 \Upsilon_1 \Upsilon_2 \Upsilon_3 \biggl(
  \frac{1}{2 \qcyl^2} \Upsilon_1 \Upsilon_2 \Upsilon_3\\
    - |\kgeo_1|^2 \Upsilon_2 \Upsilon_3
       - \Upsilon_1 |\kgeo_2|^2 \Upsilon_3 - \Upsilon_1 \Upsilon_2 |\kgeo_3|^2 \biggr) = 0
\label{eq:factorized.determinant}
\end{multline}
at a given rational surface $\zeta(\qnot) = 0$ labelled by the safety factor
$\qnot = -m/n$, with $\Upsilon_p = (\wtilde / \wS)^2 - p^2$ and
\begin{equation}
\wS^2 =  \frac{\btilde}{\rhotilde \qnot^2 (1 + \btilde)}.
\label{eq:omegaS}
\end{equation}
Aside from the trivial solution $\wtilde = 0$, the roots $\Upsilon_p = 0$ for $p
= 1,2,3$ are the top branches of the three gaps located where the cylindrical
continua of $\xiStilde_{m \pm p}$ would cross each other
(Fig.~\ref{fig:simplified.continua}). At integer multiples of $\wS$, these roots
are independent of the equilibrium shaping and the first one corresponds to the
well-known frequency $\wBAAE$~\cite{holst.2000b, gorelenkov.2007,
gorelenkov.2007b}. In turn, the factor in brackets yields three more roots: the
lower branches of the $p = 2,3$ gaps at
\begin{equation}
\wtilde^2 \big/ \wS^2 =
  p^2 - \bigl( p^2 - 1 \bigr) C_p |\kgeo_p|^2_2 + \cdots
\label{eq:bottom.branch.resonant.surface}
\end{equation}
and the bottom of the SA continuum that is uplifted to the geodesic frequency
$\wG$ defined as
\begin{equation}
\frac{\wG^2}{\wS^2} = 1 + 2 \qnot^2 \biggl(
  1 + |\kgeo_1|^2_1 + \sum_{p = 1}^3 C_p |\kgeo_p|^2_2 + \cdots \biggr),
\label{eq:geodesic.frequency}
\end{equation}
with $1/C_p = 1 - \bigl( p^2 - 1 \bigr) \big/ \bigl( 2 \qnot^2 \bigr)$, whereas
$|\kgeo_1|^2_1 = -\tfrac{3}{2} \dtilde \sym{\kappa}$ and $|\kgeo_p|^2_2$ are the
linear and quadratic terms of $|\kgeo_p|^2$, all of which depend on the shaping
as follows from Eqs.~\eqref{eq:curvature.coefficients}. In the limit of circular
equilibria $|\kgeo_1|^2_1$ and $|\kgeo_p|^2_2$ vanish, thus reducing the
frequency $\bigl(\vA^0/R_0\bigr) \wG$ to $\wGAM$ as obtained in previous
works~\cite{holst.2000b, gorelenkov.2007, gorelenkov.2007b, cheng.2019} and
actually closing the $p = 2,3$ gaps. Because their width is 
$\btilde^\frac{1}{2} O(\etilde^2 \sym{\Delta}^2, \dtilde^2 \sym{\kappa}^2)$,
such gaps are only relevant for large values of $\btilde$, $\sym{\Delta}$, or
$\sym{\kappa}$.

Slightly away from the rational surface $\zeta(\qnot) = 0$, wider gaps of size
$\btilde^\frac{1}{2} O(\etilde \sym{\Delta}, \dtilde \sym{\kappa})$ arise if the
$\xiStilde_{m\pm p}$ continuum crosses the one from $\xiAtilde_m$
(Fig.~\ref{fig:simplified.continua}), whose bottom is uplifted to $\wG$. The
condition enabling such crossings is therefore
\begin{equation}
\wG^2 < p^2 \wS^2 \quad \Leftrightarrow \quad
\qnot^2 < \frac{1}{2} \frac{p^2 - 1}{
  1 - \tfrac{3}{2} \dtilde \sym{\kappa}} + \cdots
\label{eq:hogae.condition}
\end{equation}
and gaps with $|p| = 2,3$ may open if $\qnot \lesssim \sqrt{3/2}$ or $2$,
respectively, with elongation shifting these limits slightly upwards. The locus
$\zeta_p$ of such gaps is found replacing $\wtilde^2 = \wS^2 \qnot^2 (\zeta +
p/\qcyl)^2$ from Eq.~\eqref{eq:cylindrical.continua} and $\qcyl = \qnot/(1 -
\zeta/n)$ into the submatrix $\mathsf{C}_{\pm 1}(\wtilde^2, \zeta)$ obtained
from $\mathsf{C}(\wtilde^2, \zeta)$ by keeping only the harmonics $\xiAtilde_m$
and $\xiStilde_{m \pm 1}$ in Eq.~\eqref{eq:coupling.matrix}. The condition $\det
\mathsf{C}_{\pm 1}(\zeta) = 0$ is solved assuming $\zeta$ to be a series in the
small parameter $\varsigma = \btilde \big/ \bigl[ \qnot^2 (1 - \wS^2/\wG^2)
\bigr]$, yielding at length for each $p$
\begin{equation}
\zeta_p = \pm \varsigma^\frac{1}{2} \sqrt{p^2 - \wG^2/\wS^2}
  + \varsigma \bigl(1 + n p \qnot - p^2\bigr)/n + \cdots.
\label{eq:hogae.zeta}
\end{equation}
Requiring a real-valued $\zeta_p$ above recovers Eq.~\eqref{eq:hogae.condition},
while the condition $n \qcyl + m = \qcyl \zeta_p$ becomes the analog of the well
known rule $n \qcyl + m = p/2$ that is valid for shape-induced couplings of SA
continua~\cite{heidbrink.2008}.

\begin{figure}
\includegraphics[width=246pt]{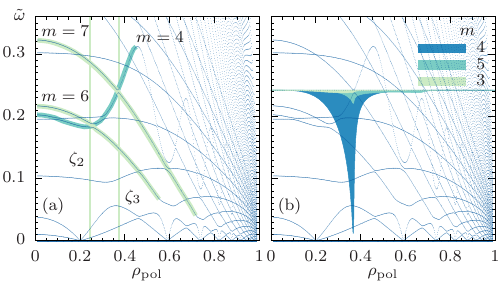}
\caption{\label{fig:jetlike.continua}
Continua for the JET-like equilibrium and $n = -3$: (a) coupling of the
$\xiAtilde_4$, $\xiStilde_6$, and $\xiStilde_7$ branches ($p = 2,3$), their
gaps, and locations by $\zeta_p$; (b) global eigenmode ($q B_\phi^{-1} \vbf{\xi}
\cdot \nabla \rhopol^2$ in a.u., dominant harmonics only) in the $p = 3$ gap.}
\end{figure}

Continua for tokamak equilibria keep the key features discussed above. The
numerical results of a continuous-spectrum extension~\cite{poedts.1993} to the
MHD code \texttt{CASTOR}~\cite{kerner.1998} are plotted in
Fig.~\ref{fig:jetlike.continua}, for the JET-like equilibrium of
Fig.~\ref{fig:jetlike.equilibrium}. Using data from the latter, one finds the
values $\btilde \approx 0.017$ and $\wS \approx 0.1$ over the rational surface
$\qnot = 4/3$ located at $\rhopol = 0.2$, while keeping quadratic terms in
Eq.~\eqref{eq:geodesic.frequency} yields $\wG \approx 0.18$. All these values
agree with the plotted results. Besides the tiny gaps at $2 \wS$ and $3 \wS$,
the local value of the ratio $\wG^2/\wS^2 \approx 3.24$ in
Eq.~\eqref{eq:hogae.condition} predicts the wider $p = 2,3$ gaps, which are
clearly visible in Fig.~\ref{fig:jetlike.continua} slightly away from the
rational surface. Their locations are provided by Eq.~\eqref{eq:hogae.zeta},
after the estimates $\zeta_2$ and $\zeta_3$ are translated into safety-factor
values and then converted to radial positions using the profile $q(\rhopol)$ in
Fig.~\ref{fig:jetlike.equilibrium}. These locations correspond to the vertical
lines in Fig.~\ref{fig:jetlike.continua}~(a), again in agreement with the
plotted numerical spectrum.

Inside these high-order frequency gaps (i.e., $|p| = 2,3$ due to equilibrium
shaping as opposed to $|p| = 1$ caused by low-order toroidicity), traveling
waves are replaced by high-order geodesic-acoustic eigenmodes (HOGAEs), as the
one computed by \texttt{CASTOR} and depicted in
Fig.~\ref{fig:jetlike.continua}~(b). Replacing Eq.~\eqref{eq:hogae.zeta} in the
acoustic continuum of Eq.~\eqref{eq:cylindrical.continua}, their frequency is
\begin{equation}
\wtilde_p/\wS = \bigl| p \bigr| \pm \varsigma^\frac{1}{2} \qnot
  \bigr(1 + p/m\bigr) \sqrt{p^2 - \wG^2/\wS^2} + \cdots
\label{eq:hogae.frequency}
\end{equation}
and the estimate $\wtilde_3 \approx 0.235$ agrees well with the plotted value.
Recalling that $\tilde{\omega}_\text{TAE} = 1/(2\qcyl)$, one finds at lowest
order from Eq.~\eqref{eq:hogae.frequency} the ratio
$\wtilde_p/\tilde{\omega}_\text{TAE} \approx 2 p \btilde^\frac{1}{2} \sim p/5$,
if $\btilde \sim 10^{-2}$ as is usually the case in tokamaks. HOGAEs with $|p| =
3$ are of particular interest because, at the same location, their frequency
$\tilde{\omega}_3 \sim \tfrac{3}{5} \tilde{\omega}_\text{TAE}$ is the one
closest to the TAE gap. In fact, the value $\tilde{\omega}_3$ evaluated at the
plasma core can be larger than $\tilde{\omega}_\text{TAE}$ for outer TAEs
located at higher $q$. Therefore, HOGAEs may play a role similar to that of TAEs
in the stability of fusion plasmas.

\section{
\label{sec:resonant.interactions}
Resonant interactions and linear stability}

\begin{figure}
\includegraphics[width=246pt]{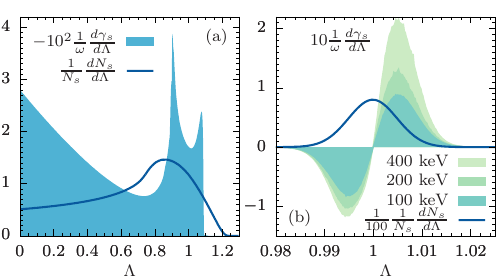}
\caption{\label{fig:lambda.spectra}
Normalised linear growth rate $\species{\gamma}/\omega$ and number of particles
$\species{N}$ per $\Lambda$ unit for thermal (a) and ICRH ions (b).}
\end{figure}

The interaction between HOGAEs and a species $s$ can be evaluated
perturbatively~\cite{porcelli.1994} if the current density $J_s$ follows the
condition
\begin{equation}
\frac{\species{J}}{J} \sim \species{Z} \frac{n_s}{n_\text{e}}
  \biggl( \frac{m_\text{e}}{\species{m}} \biggr)^\frac{1}{2}
    \biggl( \frac{\species{T}}{T_\text{e}} \biggr)^\frac{1}{2} \ll 1
\end{equation}
(with $\species{Z}$, $\species{m}$, $\species{n}$, and $\species{T}$ the charge
number, mass, particle density, and temperature) and if the growth rate
$\species{\gamma}$ is such that
\begin{equation}
\frac{\species{\gamma}}{\omega} = -\text{Im} \int
  \frac{L_{(1)}^\ast \species{f}^{(1)}}{2 \omega^2} d^3x d^3v
  \bigg/ \int \rho \vbf{\xi} \cdot \vbf{\xi}^\ast d^3x \ll 1.
\label{eq:growth.rate}
\end{equation}
Here, $L_{(1)}$ and $\species{f}^{(1)}$ are the linear response of the
guiding-center Lagrangian and equilibrium distribution function $\species{f}$ to
the perturbation $\vbf{\xi}$~\cite{porcelli.1994}. The integrals in
Eq.~\eqref{eq:growth.rate} are computed by the drift-kinetic code
\texttt{CASTOR-K}~\cite{borba.1999, nabais.2015} in the space of the
guiding-center constants of motion: energy $E$, toroidal momentum $P_\phi$, and
$\Lambda = \mu B_0/E$, with $\mu$ the magnetic moment. The results for the
interaction between the $p = 3$ HOGAE in Fig.~\ref{fig:jetlike.continua},
thermal deuterium (D, Maxwellian distribution with $T_\text{D} = T_\text{e}$ as
in Fig.~\ref{fig:jetlike.equilibrium}) and ion-cyclotron resonance heating
(ICRH) H ions are shown in Fig.~\ref{fig:lambda.spectra}, assuming a separable
distribution
\begin{equation}
f_\text{ICRH}(\rhopol, E, \Lambda) \propto
  \frac{1 - \rhopol}{\sqrt{0.015 + \rhopol}} \,
    e^{-\frac{E}{T_\text{H}}} \,
      e^{-\frac{(\Lambda - 1)^2}{2 \delta_\Lambda^2}},
\label{eq:icrh.distro}
\end{equation}
with $\delta_\Lambda = 1/200$ corresponding to a Doppler broadening $2 R_0
\delta_\Lambda = 3$~cm of the ICRH resonant layer. Damping on thermal ions is
mainly due to passing particles ($\Lambda \lesssim 1 - \varepsilon$) and
$\gamma_\text{D}/\omega = -0.0175$. Trapped ICRH ions produce drive and damping
at $\Lambda = 1 \pm \delta_\Lambda$ respectively, where $|\partial f_s/
\partial\Lambda |$ is highest. However, the energy transfer is larger for higher
$\Lambda$, whence a net drive that increases with $T_\text{H}$. The growth rate
$\gamma_\text{H}/\omega$ is listed in Tab.~\ref{tab:icrh.gammas} for different
$T_\text{H}$ values and the instability threshold is thus slightly above
$200$~keV.
\begin{table}
\caption{\label{tab:icrh.gammas}
Normalised growth rate due to ICRH accelerated H ions
assuming a particle-number ratio $N_\text{H}/N_\text{D} = \tfrac{1}{100}$.}
\begin{ruledtabular}
\begin{tabular}{r|rrrr}
$T_\text{H}$ (keV) & $100$ & $200$ & $400$ & $800$ \\
\hline
$\gamma_\text{H}/\omega$ & $0.004$ & $0.012$ & $0.049$ & $0.089$
\end{tabular}
\end{ruledtabular}
\end{table}

\begin{figure}
\includegraphics[width=246pt]{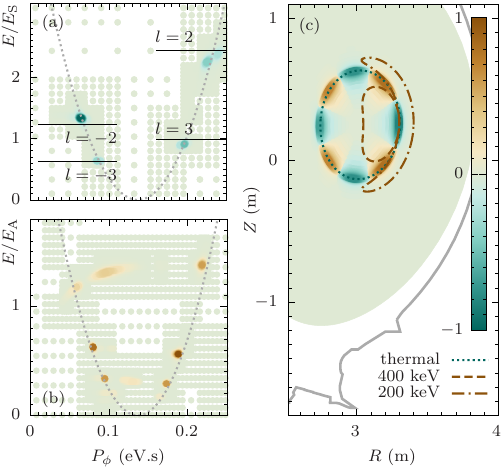}
\caption{\label{fig:resonances}
Energy exchange (shading code, a.u.) due to passing ($\Lambda = 10^{-3}$)
thermal (a) and trapped ($\Lambda = 1 + \delta_\Lambda$) ICRH ions (b,
$T_\text{H} = 200$~keV), along with the gap radial location (dotted line); HOGAE
poloidal structure ($q B_\phi^{-1} \vbf{\xi} \cdot \nabla \rhopol^2$ in a.u.)
and orbits with largest energy transfer (c).}
\end{figure}

For resonant interactions between AEs and particles to take place, $\omega$ must
be related with the orbit-averaged frequencies $\langle \dot{\theta} \rangle$
and
$\langle \dot{\phi} \rangle$ as
\begin{equation}
\omega + n \langle \dot{\phi} \rangle +
  \bigl(l + m\bigr) \langle \dot{\theta} \rangle = 0
\label{eq:resonance.condition}
\end{equation}
with $l$ an integer~\cite{heidbrink.2008}. In the strongly passing-particle
limit $\Lambda \rightarrow 0$, these are $\langle \dot{\phi} \rangle \approx q
\langle \dot{\theta} \rangle \approx v_\parallel/R_0$ and
Eq.~\eqref{eq:resonance.condition} becomes $\wtilde + \bigl( \zeta_p + l/q
\bigr) \bigl( v_\parallel/\vA^0 \bigr) = 0$. Replacing $\wtilde$ and $\zeta_p$
by Eqs.~\eqref{eq:hogae.frequency} and~\eqref{eq:hogae.zeta} and solving for
$v_\parallel$ yields the series
\begin{equation}
\frac{v_\parallel}{\cS} = -\frac{\bigl| p \bigr|}{l \sqrt{1 + \btilde}} \Biggl[
  1 + \varsigma^\frac{1}{2} \qnot C_{l,m,p}
    \sqrt{ p^2 - \frac{\wG^2}{\wS^2} } + \cdots
      \Biggr],
\label{eq:passing.resonances}
\end{equation}
with $C_{l,m,p} = \pm (m + p)/(m |p|) \mp (l + m)/(l m)$, where the top/bottom
choices for the signs correspond to those made for the gap frequency and
location in Eqs.~\eqref{eq:hogae.zeta} and~\eqref{eq:hogae.frequency},
respectively. The condition in Eq.~\eqref{eq:passing.resonances} is the
equivalent to the known relation $v_\parallel/\vA^0 = - p'/(2l + 1)$ for SA
eigenmodes (and, in particular, for TAEs when $p' = 1$)~\cite{heidbrink.2008}.
The interaction of thermal and ICRH ions with the $p = 3$ HOGAE in the $\{ E,
P_\phi \}$ plane is displayed in Fig.~\ref{fig:resonances}, with $E_\text{A} =
\tfrac{1}{2} \species{m} \vA^2$ and $E_\text{S} = \tfrac{1}{2} \species{m}
\cS^2$. Strongly-passing thermal ions show resonances along the gap radial
location, at energy values in agreement with the estimates produced by
Eq.~\eqref{eq:passing.resonances} for several values of the integer number $l$.
Their temperature ($T_\text{D} \approx 5$ keV) is not sufficient to access the
fundamental resonances ($|l| = 1$, at $128.1$ keV and $31.2$~keV) and
interactions are restricted to lower sidebands ($|l| \geqslant 2$). By their
side, trapped ICRH-ion resonances depend on bounce and precession frequencies
and their interaction pattern is thus more complex.

\begin{figure}
\includegraphics[width=246pt]{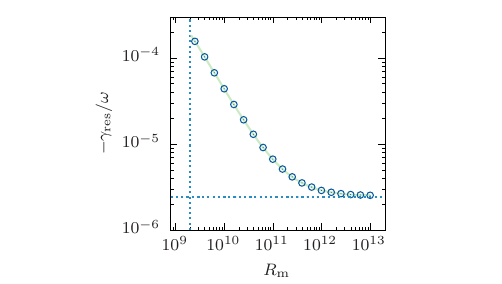}
\caption{\label{fig:damping}
Normalised damping rate $\gammares/\omega$ from resistive MHD as a function of
the magnetic Reynolds number $\Rm$ for the $n = -3, p = 3$ HOGAE.}
\end{figure}

Besides the thermal-ion Landau damping discussed in the previous paragraphs,
other damping mechanisms are usually taken into account when assessing the
stability of AEs in fusion devices. These include the collisional damping on
trapped electrons~\cite{gorelenkov.1992}, the radiative damping due to finite
coupling with kinetic \alfven{} waves~\cite{mett.1992, candy.1994}, and the
damping caused by eventual interactions with the ideal-MHD
continuum~\cite{rosenbluth.1992, zonca.1992}. All these damping models, however,
were developed with SA waves in mind (particularly TAEs) and, in their present
form, are not suitable to handle AEs with a non-neglegible acoustic component,
as is the case of HOGAEs. One noticeable exception is the evaluation of the
continuum damping by taking the imaginary part of the eigenvalue $\omega + i
\gammares$ of the linear resistive-MHD problem in the limit of vanishing plasma
resistivity~\cite{poedts.1991, poedts.1992}. This aim can be accomplished with
the MHD code \texttt{CASTOR} (keeping compressibility and finite resistivity
$\eta$) and Fig.~\ref{fig:damping} displays a scan in the magnetic Reynolds
number $\Rm = \mu_0 \vA^0 R_0/\eta$ that starts at the value of the Spitzer
resistivity (corresponding to $1/\Rm \approx 5 \times 10^{-10}$) and spans about
four orders of magnitude. As expected, the normalised damping rate becomes
independent of the resistivity for large $\Rm$~\cite{poedts.1991, poedts.1992}.
Moreover, the asymptotic value $\gammares/\omega \approx 2.4 \times 10^{-6}$
indicates a very weak interaction between the ideal-MHD continuum and the
considered HOGAE.

\section{
\label{sec:discussion}
Discussion}

In summary, equilibrium shaping (mainly due to $\sym{\Delta}$ and
$\sym{\kappa}$) was shown to couple acoustic and SA continua through $p$-order
periodicity in $\kgeo(\stilde,\theta)$, leading to high-order (i.e., $|p| =
2,3)$ frequency gaps at $\wGAM \lesssim \omega_p \lesssim \omega_\text{TAE}$
that lie significantly above the previously known toroidicity-induced BAAE gap
($|p| = 1$), if the condition in Eq.~\eqref{eq:hogae.condition} is met. Inside
such gaps, global HOGAEs were found to be driven unstable by anisotropic
ICRH-ion populations for tokamak parameters. Circulating-ion resonances were
shown to lie in the range $E \lesssim p^2 E_S$, this limit being near the
geometric mean $(E_\text{S} E_\text{A})^\frac{1}{2}$ because $E_\text{S} /
E_\text{A} \sim \btilde$ and $p^2 \btilde^\frac{1}{2} \sim 1$ in fusion devices.

In conclusion, potentially unstable HOGAEs were found to populate the frequency
range below but close to $\omega_\text{TAE}$ in tokamak plasmas.  Their location
and frequency estimates in Eqs.~\eqref{eq:hogae.zeta} and
\eqref{eq:hogae.frequency} are expected to guide the interpretation of their
eventual observation in experiments~\cite{rodrigues.2021fec} and their role in
simulations of SA-acoustic continua coupling for next-step fusion
devices~\cite{todo.2014, bierwage.2015}, as well as to foster their use in MHD
spectroscopy.

Among all AEs induced by finite $\beta$ (i.e., ASEs), as those found in many
numerical approaches~\cite{huysmans.1995, cheng.2019, kramer.2020}, HOGAEs have
the frequency closer to $\wTAE$ and, therefore, the possibility to interact with
energetic ions in a fashion similar to that of TAEs. Indeed, the resonance
condition in Eq.~\eqref{eq:passing.resonances} shows that HOGAEs are able to tap
energy from hot ions at $E \sim \btilde^\frac{1}{2} E_\text{A}$, while avoiding
efficient damping by thermal ions at the fundamental resonance $E \sim p^2
E_\text{S} \sim \gamma p^2 T_\text{D}$. Driven by less energetic ions, HOGAEs
are thus expected to grow slower than TAEs that have resonant interactions at $E
\sim E_\text{A}$. Yet, they may dominate at the nonlinear saturation stage, as
found numerically for low-frequency AEs close to $\wGAM$~\cite{todo.2014}.
HOGAEs may hence play a significant and still unexplored role, along with TAEs,
in the stability assessments of fusion reactors like ITER~\cite{pinches.2015,
lauber.2015, rodrigues.2015, figueiredo.2016, fitzgerald.2016, schneller.2016}.
Three issues beyond the scope of this work should be pursued elsewhere:
characterise HOGAEs in experimental scenarios as those recently
reported~\cite{rodrigues.2021fec}, evaluate their interaction with isotropic
$\alpha$-particles, and estimate the hot-ion redistribution and losses they may
induce in burning plasmas.

\begin{acknowledgments}
The authors gratefully thank Drs.~S.~Sharapov and M.~Fitzgerald (CCFE, UK) for
insightful discussions. IPFN activities were supported by Funda\c{c}\~{a}o para
a Ci\^{e}ncia e Tecnologia (FCT, Lisboa) via project UID/FIS/50010/2019. One of
the authors (FC) was supported by FuseNet, Euratom's research and training
programme in the EUROfusion Consortium, under Grant No.~633053. Views and
opinions expressed here do not necessarily reflect those of the European
Commission.
\end{acknowledgments}


\begin{thebibliography}{61}%
\makeatletter
\providecommand \@ifxundefined [1]{%
 \@ifx{#1\undefined}
}%
\providecommand \@ifnum [1]{%
 \ifnum #1\expandafter \@firstoftwo
 \else \expandafter \@secondoftwo
 \fi
}%
\providecommand \@ifx [1]{%
 \ifx #1\expandafter \@firstoftwo
 \else \expandafter \@secondoftwo
 \fi
}%
\providecommand \natexlab [1]{#1}%
\providecommand \enquote  [1]{``#1''}%
\providecommand \bibnamefont  [1]{#1}%
\providecommand \bibfnamefont [1]{#1}%
\providecommand \citenamefont [1]{#1}%
\providecommand \href@noop [0]{\@secondoftwo}%
\providecommand \href [0]{\begingroup \@sanitize@url \@href}%
\providecommand \@href[1]{\@@startlink{#1}\@@href}%
\providecommand \@@href[1]{\endgroup#1\@@endlink}%
\providecommand \@sanitize@url [0]{\catcode `\\12\catcode `\$12\catcode
  `\&12\catcode `\#12\catcode `\^12\catcode `\_12\catcode `\%12\relax}%
\providecommand \@@startlink[1]{}%
\providecommand \@@endlink[0]{}%
\providecommand \url  [0]{\begingroup\@sanitize@url \@url }%
\providecommand \@url [1]{\endgroup\@href {#1}{\urlprefix }}%
\providecommand \urlprefix  [0]{URL }%
\providecommand \Eprint [0]{\href }%
\providecommand \doibase [0]{http://dx.doi.org/}%
\providecommand \selectlanguage [0]{\@gobble}%
\providecommand \bibinfo  [0]{\@secondoftwo}%
\providecommand \bibfield  [0]{\@secondoftwo}%
\providecommand \translation [1]{[#1]}%
\providecommand \BibitemOpen [0]{}%
\providecommand \bibitemStop [0]{}%
\providecommand \bibitemNoStop [0]{.\EOS\space}%
\providecommand \EOS [0]{\spacefactor3000\relax}%
\providecommand \BibitemShut  [1]{\csname bibitem#1\endcsname}%
\let\auto@bib@innerbib\@empty
\bibitem [{\citenamefont {Uberoi}(1972)}]{uberoi.1972}%
  \BibitemOpen
  \bibfield  {author} {\bibinfo {author} {\bibfnamefont {C.}~\bibnamefont
  {Uberoi}},\ }\href {\doibase 10.1063/1.1694148} {\bibfield  {journal}
  {\bibinfo  {journal} {Phys. Fluids}\ }\textbf {\bibinfo {volume} {15}},\
  \bibinfo {pages} {1673} (\bibinfo {year} {1972})}\BibitemShut {NoStop}%
\bibitem [{\citenamefont {Grad}(1973)}]{grad.1973}%
  \BibitemOpen
  \bibfield  {author} {\bibinfo {author} {\bibfnamefont {H.}~\bibnamefont
  {Grad}},\ }\href {\doibase 10.1073/pnas.70.12.3277} {\bibfield  {journal}
  {\bibinfo  {journal} {Proc. Natl. Acad. Sci. USA}\ }\textbf {\bibinfo
  {volume} {70}},\ \bibinfo {pages} {3277} (\bibinfo {year}
  {1973})}\BibitemShut {NoStop}%
\bibitem [{\citenamefont {Goedbloed}(1975)}]{goedbloed.1975}%
  \BibitemOpen
  \bibfield  {author} {\bibinfo {author} {\bibfnamefont {J.~P.}\ \bibnamefont
  {Goedbloed}},\ }\href {\doibase 10.1063/1.861012} {\bibfield  {journal}
  {\bibinfo  {journal} {Phys. Fluids}\ }\textbf {\bibinfo {volume} {18}},\
  \bibinfo {pages} {1258} (\bibinfo {year} {1975})}\BibitemShut {NoStop}%
\bibitem [{\citenamefont {Bernstein}\ \emph {et~al.}(1958)\citenamefont
  {Bernstein}, \citenamefont {Frieman}, \citenamefont {Kruskal},\ and\
  \citenamefont {Kulsrud}}]{bernstein.1958}%
  \BibitemOpen
  \bibfield  {author} {\bibinfo {author} {\bibfnamefont {I.~B.}\ \bibnamefont
  {Bernstein}}, \bibinfo {author} {\bibfnamefont {E.~A.}\ \bibnamefont
  {Frieman}}, \bibinfo {author} {\bibfnamefont {M.~D.}\ \bibnamefont
  {Kruskal}}, \ and\ \bibinfo {author} {\bibfnamefont {R.~M.}\ \bibnamefont
  {Kulsrud}},\ }\href@noop {} {\bibfield  {journal} {\bibinfo  {journal} {Proc.
  Roy. Soc. Series A}\ }\textbf {\bibinfo {volume} {244}},\ \bibinfo {pages}
  {17} (\bibinfo {year} {1958})}\BibitemShut {NoStop}%
\bibitem [{\citenamefont {Abramowitz}\ and\ \citenamefont
  {Stegun}(1972)}]{abramowitz.1972}%
  \BibitemOpen
  \bibfield  {author} {\bibinfo {author} {\bibfnamefont {M.}~\bibnamefont
  {Abramowitz}}\ and\ \bibinfo {author} {\bibfnamefont {I.~A.}\ \bibnamefont
  {Stegun}},\ }\href@noop {} {\emph {\bibinfo {title} {Handbook of Mathematical
  Functions}}},\ \bibinfo {edition} {9th}\ ed.\ (\bibinfo  {publisher}
  {Dover},\ \bibinfo {year} {1972})\BibitemShut {NoStop}%
\bibitem [{\citenamefont {Tataronis}\ and\ \citenamefont
  {Grossmann}(1973)}]{tataronis.1973}%
  \BibitemOpen
  \bibfield  {author} {\bibinfo {author} {\bibfnamefont {J.}~\bibnamefont
  {Tataronis}}\ and\ \bibinfo {author} {\bibfnamefont {W.}~\bibnamefont
  {Grossmann}},\ }\href {\doibase 10.1007/BF01391913} {\bibfield  {journal}
  {\bibinfo  {journal} {Z. Physik}\ }\textbf {\bibinfo {volume} {261}},\
  \bibinfo {pages} {203} (\bibinfo {year} {1973})}\BibitemShut {NoStop}%
\bibitem [{\citenamefont {Grossmann}\ and\ \citenamefont
  {Tataronis}(1973)}]{grossmann.1973}%
  \BibitemOpen
  \bibfield  {author} {\bibinfo {author} {\bibfnamefont {W.}~\bibnamefont
  {Grossmann}}\ and\ \bibinfo {author} {\bibfnamefont {J.}~\bibnamefont
  {Tataronis}},\ }\href {\doibase 10.1007/BF01391914} {\bibfield  {journal}
  {\bibinfo  {journal} {Z. Physik}\ }\textbf {\bibinfo {volume} {261}},\
  \bibinfo {pages} {217} (\bibinfo {year} {1973})}\BibitemShut {NoStop}%
\bibitem [{\citenamefont {Appert}\ \emph {et~al.}(1974)\citenamefont {Appert},
  \citenamefont {Gruber},\ and\ \citenamefont {Vaclavik}}]{appert.1974}%
  \BibitemOpen
  \bibfield  {author} {\bibinfo {author} {\bibfnamefont {K.}~\bibnamefont
  {Appert}}, \bibinfo {author} {\bibfnamefont {R.}~\bibnamefont {Gruber}}, \
  and\ \bibinfo {author} {\bibfnamefont {J.}~\bibnamefont {Vaclavik}},\ }\href
  {\doibase 10.1063/1.1694918} {\bibfield  {journal} {\bibinfo  {journal}
  {Phys. Fluids}\ }\textbf {\bibinfo {volume} {17}},\ \bibinfo {pages} {1471}
  (\bibinfo {year} {1974})}\BibitemShut {NoStop}%
\bibitem [{\citenamefont {Strutt}(1887)}]{rayleigh.1887}%
  \BibitemOpen
  \bibfield  {author} {\bibinfo {author} {\bibfnamefont {J.~W.}\ \bibnamefont
  {Strutt}},\ }\href {\doibase 10.1080/14786448708628074} {\bibfield  {journal}
  {\bibinfo  {journal} {Phil. Mag.}\ }\textbf {\bibinfo {volume} {24}},\
  \bibinfo {pages} {145} (\bibinfo {year} {1887})}\BibitemShut {NoStop}%
\bibitem [{\citenamefont {Cheng}\ and\ \citenamefont
  {Chance}(1986)}]{cheng.1986}%
  \BibitemOpen
  \bibfield  {author} {\bibinfo {author} {\bibfnamefont {C.~Z.}\ \bibnamefont
  {Cheng}}\ and\ \bibinfo {author} {\bibfnamefont {M.~S.}\ \bibnamefont
  {Chance}},\ }\href {\doibase 10.1063/1.865801} {\bibfield  {journal}
  {\bibinfo  {journal} {Phys. Fluids}\ }\textbf {\bibinfo {volume} {29}},\
  \bibinfo {pages} {3695} (\bibinfo {year} {1986})}\BibitemShut {NoStop}%
\bibitem [{\citenamefont {Zhang}\ \emph {et~al.}(2008)\citenamefont {Zhang},
  \citenamefont {Heidbrink}, \citenamefont {Boehmer}, \citenamefont
  {McWilliams}, \citenamefont {Chen}, \citenamefont {Breizman}, \citenamefont
  {Vincena}, \citenamefont {Carter}, \citenamefont {Leneman}, \citenamefont
  {Gekelman}, \citenamefont {Pribyl},\ and\ \citenamefont
  {Brugman}}]{zhang.2008}%
  \BibitemOpen
  \bibfield  {author} {\bibinfo {author} {\bibfnamefont {Y.}~\bibnamefont
  {Zhang}}, \bibinfo {author} {\bibfnamefont {W.~W.}\ \bibnamefont
  {Heidbrink}}, \bibinfo {author} {\bibfnamefont {H.}~\bibnamefont {Boehmer}},
  \bibinfo {author} {\bibfnamefont {R.}~\bibnamefont {McWilliams}}, \bibinfo
  {author} {\bibfnamefont {G.}~\bibnamefont {Chen}}, \bibinfo {author}
  {\bibfnamefont {B.~N.}\ \bibnamefont {Breizman}}, \bibinfo {author}
  {\bibfnamefont {S.}~\bibnamefont {Vincena}}, \bibinfo {author} {\bibfnamefont
  {T.}~\bibnamefont {Carter}}, \bibinfo {author} {\bibfnamefont
  {D.}~\bibnamefont {Leneman}}, \bibinfo {author} {\bibfnamefont
  {W.}~\bibnamefont {Gekelman}}, \bibinfo {author} {\bibfnamefont
  {P.}~\bibnamefont {Pribyl}}, \ and\ \bibinfo {author} {\bibfnamefont
  {B.}~\bibnamefont {Brugman}},\ }\href {\doibase 10.1063/1.2827518} {\bibfield
   {journal} {\bibinfo  {journal} {Phys. Plasmas}\ }\textbf {\bibinfo {volume}
  {15}},\ \bibinfo {pages} {012103} (\bibinfo {year} {2008})}\BibitemShut
  {NoStop}%
\bibitem [{\citenamefont {Rosenbluth}\ and\ \citenamefont
  {Rutherford}(1975)}]{rosenbluth.1975}%
  \BibitemOpen
  \bibfield  {author} {\bibinfo {author} {\bibfnamefont {M.~N.}\ \bibnamefont
  {Rosenbluth}}\ and\ \bibinfo {author} {\bibfnamefont {P.~H.}\ \bibnamefont
  {Rutherford}},\ }\href {\doibase 10.1103/PhysRevLett.34.1428} {\bibfield
  {journal} {\bibinfo  {journal} {Phys. Rev. Lett.}\ }\textbf {\bibinfo
  {volume} {34}},\ \bibinfo {pages} {1428} (\bibinfo {year}
  {1975})}\BibitemShut {NoStop}%
\bibitem [{\citenamefont {Fu}\ and\ \citenamefont {Dam}(1989)}]{fu.1989}%
  \BibitemOpen
  \bibfield  {author} {\bibinfo {author} {\bibfnamefont {G.~Y.}\ \bibnamefont
  {Fu}}\ and\ \bibinfo {author} {\bibfnamefont {J.~W.~V.}\ \bibnamefont
  {Dam}},\ }\href {\doibase 10.1063/1.859057} {\bibfield  {journal} {\bibinfo
  {journal} {Phys. Fluids B}\ }\textbf {\bibinfo {volume} {1}},\ \bibinfo
  {pages} {1949} (\bibinfo {year} {1989})}\BibitemShut {NoStop}%
\bibitem [{\citenamefont {Betti}\ and\ \citenamefont
  {Freidberg}(1992)}]{betti.1992}%
  \BibitemOpen
  \bibfield  {author} {\bibinfo {author} {\bibfnamefont {R.}~\bibnamefont
  {Betti}}\ and\ \bibinfo {author} {\bibfnamefont {J.~P.}\ \bibnamefont
  {Freidberg}},\ }\href {\doibase 10.1063/1.860057} {\bibfield  {journal}
  {\bibinfo  {journal} {Phys. Fluids B}\ }\textbf {\bibinfo {volume} {4}},\
  \bibinfo {pages} {1465} (\bibinfo {year} {1992})}\BibitemShut {NoStop}%
\bibitem [{\citenamefont {Fasoli}\ \emph {et~al.}(2007)\citenamefont {Fasoli},
  \citenamefont {Gormenzano}, \citenamefont {Berk}, \citenamefont {Breizman},
  \citenamefont {Briguglio}, \citenamefont {Darrow}, \citenamefont
  {Gorelenkov}, \citenamefont {Heidbrink}, \citenamefont {Jaun}, \citenamefont
  {Konovalov}, \citenamefont {Nazikian}, \citenamefont {Noterdaeme},
  \citenamefont {Sharapov}, \citenamefont {Shinohara}, \citenamefont {Testa},
  \citenamefont {Tobita}, \citenamefont {Todo}, \citenamefont {Vlad},\ and\
  \citenamefont {Zonca}}]{fasoli.2007}%
  \BibitemOpen
  \bibfield  {author} {\bibinfo {author} {\bibfnamefont {A.}~\bibnamefont
  {Fasoli}}, \bibinfo {author} {\bibfnamefont {C.}~\bibnamefont {Gormenzano}},
  \bibinfo {author} {\bibfnamefont {H.}~\bibnamefont {Berk}}, \bibinfo {author}
  {\bibfnamefont {B.}~\bibnamefont {Breizman}}, \bibinfo {author}
  {\bibfnamefont {S.}~\bibnamefont {Briguglio}}, \bibinfo {author}
  {\bibfnamefont {D.}~\bibnamefont {Darrow}}, \bibinfo {author} {\bibfnamefont
  {N.}~\bibnamefont {Gorelenkov}}, \bibinfo {author} {\bibfnamefont
  {W.}~\bibnamefont {Heidbrink}}, \bibinfo {author} {\bibfnamefont
  {A.}~\bibnamefont {Jaun}}, \bibinfo {author} {\bibfnamefont {S.}~\bibnamefont
  {Konovalov}}, \bibinfo {author} {\bibfnamefont {R.}~\bibnamefont {Nazikian}},
  \bibinfo {author} {\bibfnamefont {J.-M.}\ \bibnamefont {Noterdaeme}},
  \bibinfo {author} {\bibfnamefont {S.}~\bibnamefont {Sharapov}}, \bibinfo
  {author} {\bibfnamefont {K.}~\bibnamefont {Shinohara}}, \bibinfo {author}
  {\bibfnamefont {D.}~\bibnamefont {Testa}}, \bibinfo {author} {\bibfnamefont
  {K.}~\bibnamefont {Tobita}}, \bibinfo {author} {\bibfnamefont
  {Y.}~\bibnamefont {Todo}}, \bibinfo {author} {\bibfnamefont {G.}~\bibnamefont
  {Vlad}}, \ and\ \bibinfo {author} {\bibfnamefont {F.}~\bibnamefont {Zonca}},\
  }\href {\doibase 10.1088/0029-5515/47/6/S05} {\bibfield  {journal} {\bibinfo
  {journal} {Nucl. Fusion}\ }\textbf {\bibinfo {volume} {47}},\ \bibinfo
  {pages} {S264} (\bibinfo {year} {2007})}\BibitemShut {NoStop}%
\bibitem [{\citenamefont {Heidbrink}\ and\ \citenamefont
  {Sadler}(1994)}]{heidbrink.1994}%
  \BibitemOpen
  \bibfield  {author} {\bibinfo {author} {\bibfnamefont {W.}~\bibnamefont
  {Heidbrink}}\ and\ \bibinfo {author} {\bibfnamefont {G.}~\bibnamefont
  {Sadler}},\ }\href {\doibase 10.1088/0029-5515/34/4/i07} {\bibfield
  {journal} {\bibinfo  {journal} {Nucl. Fusion}\ }\textbf {\bibinfo {volume}
  {34}},\ \bibinfo {pages} {535} (\bibinfo {year} {1994})}\BibitemShut
  {NoStop}%
\bibitem [{\citenamefont {Gorelenkov}\ \emph {et~al.}(2014)\citenamefont
  {Gorelenkov}, \citenamefont {Pinches},\ and\ \citenamefont
  {Toi}}]{gorelenkov.2014}%
  \BibitemOpen
  \bibfield  {author} {\bibinfo {author} {\bibfnamefont {N.}~\bibnamefont
  {Gorelenkov}}, \bibinfo {author} {\bibfnamefont {S.}~\bibnamefont {Pinches}},
  \ and\ \bibinfo {author} {\bibfnamefont {K.}~\bibnamefont {Toi}},\ }\href
  {\doibase 10.1088/0029-5515/54/12/125001} {\bibfield  {journal} {\bibinfo
  {journal} {Nucl. Fusion}\ }\textbf {\bibinfo {volume} {54}},\ \bibinfo
  {pages} {125001} (\bibinfo {year} {2014})}\BibitemShut {NoStop}%
\bibitem [{\citenamefont {Lauber}(2013)}]{lauber.2013}%
  \BibitemOpen
  \bibfield  {author} {\bibinfo {author} {\bibfnamefont {P.}~\bibnamefont
  {Lauber}},\ }\href {\doibase 10.1016/j.physrep.2013.07.001} {\bibfield
  {journal} {\bibinfo  {journal} {Phys. Rep.}\ }\textbf {\bibinfo {volume}
  {533}},\ \bibinfo {pages} {33 } (\bibinfo {year} {2013})}\BibitemShut
  {NoStop}%
\bibitem [{\citenamefont {Betti}\ and\ \citenamefont
  {Freidberg}(1991)}]{betti.1991}%
  \BibitemOpen
  \bibfield  {author} {\bibinfo {author} {\bibfnamefont {R.}~\bibnamefont
  {Betti}}\ and\ \bibinfo {author} {\bibfnamefont {J.~P.}\ \bibnamefont
  {Freidberg}},\ }\href {\doibase 10.1063/1.859655} {\bibfield  {journal}
  {\bibinfo  {journal} {Phys. Fluids B}\ }\textbf {\bibinfo {volume} {3}},\
  \bibinfo {pages} {1865} (\bibinfo {year} {1991})}\BibitemShut {NoStop}%
\bibitem [{\citenamefont {Heidbrink}(2008)}]{heidbrink.2008}%
  \BibitemOpen
  \bibfield  {author} {\bibinfo {author} {\bibfnamefont {W.~W.}\ \bibnamefont
  {Heidbrink}},\ }\href {\doibase 10.1063/1.2838239} {\bibfield  {journal}
  {\bibinfo  {journal} {Phys. Plasmas}\ }\textbf {\bibinfo {volume} {15}},\
  \bibinfo {pages} {055501} (\bibinfo {year} {2008})}\BibitemShut {NoStop}%
\bibitem [{\citenamefont {Heidbrink}\ \emph {et~al.}(1993)\citenamefont
  {Heidbrink}, \citenamefont {Strait}, \citenamefont {Chu},\ and\ \citenamefont
  {Turnbull}}]{heidbrink.1993}%
  \BibitemOpen
  \bibfield  {author} {\bibinfo {author} {\bibfnamefont {W.~W.}\ \bibnamefont
  {Heidbrink}}, \bibinfo {author} {\bibfnamefont {E.~J.}\ \bibnamefont
  {Strait}}, \bibinfo {author} {\bibfnamefont {M.~S.}\ \bibnamefont {Chu}}, \
  and\ \bibinfo {author} {\bibfnamefont {A.~D.}\ \bibnamefont {Turnbull}},\
  }\href {\doibase 10.1103/PhysRevLett.71.855} {\bibfield  {journal} {\bibinfo
  {journal} {Phys. Rev. Lett.}\ }\textbf {\bibinfo {volume} {71}},\ \bibinfo
  {pages} {855} (\bibinfo {year} {1993})}\BibitemShut {NoStop}%
\bibitem [{\citenamefont {Turnbull}\ \emph {et~al.}(1993)\citenamefont
  {Turnbull}, \citenamefont {Strait}, \citenamefont {Heidbrink}, \citenamefont
  {Chu}, \citenamefont {Duong}, \citenamefont {Greene}, \citenamefont {Lao},
  \citenamefont {Taylor},\ and\ \citenamefont {Thompson}}]{turnbull.1993}%
  \BibitemOpen
  \bibfield  {author} {\bibinfo {author} {\bibfnamefont {A.~D.}\ \bibnamefont
  {Turnbull}}, \bibinfo {author} {\bibfnamefont {E.~J.}\ \bibnamefont
  {Strait}}, \bibinfo {author} {\bibfnamefont {W.~W.}\ \bibnamefont
  {Heidbrink}}, \bibinfo {author} {\bibfnamefont {M.~S.}\ \bibnamefont {Chu}},
  \bibinfo {author} {\bibfnamefont {H.~H.}\ \bibnamefont {Duong}}, \bibinfo
  {author} {\bibfnamefont {J.~M.}\ \bibnamefont {Greene}}, \bibinfo {author}
  {\bibfnamefont {L.~L.}\ \bibnamefont {Lao}}, \bibinfo {author} {\bibfnamefont
  {T.~S.}\ \bibnamefont {Taylor}}, \ and\ \bibinfo {author} {\bibfnamefont
  {S.~J.}\ \bibnamefont {Thompson}},\ }\href {\doibase 10.1063/1.860742}
  {\bibfield  {journal} {\bibinfo  {journal} {Phys. Fluids B}\ }\textbf
  {\bibinfo {volume} {5}},\ \bibinfo {pages} {2546} (\bibinfo {year}
  {1993})}\BibitemShut {NoStop}%
\bibitem [{\citenamefont {Heidbrink}\ \emph {et~al.}(1999)\citenamefont
  {Heidbrink}, \citenamefont {Ruskov}, \citenamefont {Carolipio}, \citenamefont
  {Fang}, \citenamefont {van Zeeland},\ and\ \citenamefont
  {James}}]{heidbrink.1999}%
  \BibitemOpen
  \bibfield  {author} {\bibinfo {author} {\bibfnamefont {W.~W.}\ \bibnamefont
  {Heidbrink}}, \bibinfo {author} {\bibfnamefont {E.}~\bibnamefont {Ruskov}},
  \bibinfo {author} {\bibfnamefont {E.~M.}\ \bibnamefont {Carolipio}}, \bibinfo
  {author} {\bibfnamefont {J.}~\bibnamefont {Fang}}, \bibinfo {author}
  {\bibfnamefont {M.~A.}\ \bibnamefont {van Zeeland}}, \ and\ \bibinfo {author}
  {\bibfnamefont {R.~A.}\ \bibnamefont {James}},\ }\href {\doibase
  10.1063/1.873359} {\bibfield  {journal} {\bibinfo  {journal} {Phys. Plasmas}\
  }\textbf {\bibinfo {volume} {6}},\ \bibinfo {pages} {1147} (\bibinfo {year}
  {1999})}\BibitemShut {NoStop}%
\bibitem [{\citenamefont {Huysmans}\ \emph {et~al.}(1995)\citenamefont
  {Huysmans}, \citenamefont {Kerner}, \citenamefont {Borba}, \citenamefont
  {Holties},\ and\ \citenamefont {Goedbloed}}]{huysmans.1995}%
  \BibitemOpen
  \bibfield  {author} {\bibinfo {author} {\bibfnamefont {G.~T.~A.}\
  \bibnamefont {Huysmans}}, \bibinfo {author} {\bibfnamefont {W.}~\bibnamefont
  {Kerner}}, \bibinfo {author} {\bibfnamefont {D.}~\bibnamefont {Borba}},
  \bibinfo {author} {\bibfnamefont {H.~A.}\ \bibnamefont {Holties}}, \ and\
  \bibinfo {author} {\bibfnamefont {J.~P.}\ \bibnamefont {Goedbloed}},\ }\href
  {\doibase 10.1063/1.871310} {\bibfield  {journal} {\bibinfo  {journal} {Phys.
  Plasmas}\ }\textbf {\bibinfo {volume} {2}},\ \bibinfo {pages} {1605}
  (\bibinfo {year} {1995})}\BibitemShut {NoStop}%
\bibitem [{\citenamefont {Heidbrink}\ \emph {et~al.}(2021)\citenamefont
  {Heidbrink}, \citenamefont {Zeeland}, \citenamefont {Austin}, \citenamefont
  {Crocker}, \citenamefont {Du}, \citenamefont {McKee},\ and\ \citenamefont
  {Spong}}]{heidbrink.2021}%
  \BibitemOpen
  \bibfield  {author} {\bibinfo {author} {\bibfnamefont {W.}~\bibnamefont
  {Heidbrink}}, \bibinfo {author} {\bibfnamefont {M.~V.}\ \bibnamefont
  {Zeeland}}, \bibinfo {author} {\bibfnamefont {M.}~\bibnamefont {Austin}},
  \bibinfo {author} {\bibfnamefont {N.}~\bibnamefont {Crocker}}, \bibinfo
  {author} {\bibfnamefont {X.}~\bibnamefont {Du}}, \bibinfo {author}
  {\bibfnamefont {G.}~\bibnamefont {McKee}}, \ and\ \bibinfo {author}
  {\bibfnamefont {D.}~\bibnamefont {Spong}},\ }\href {\doibase
  10.1088/1741-4326/abf953} {\bibfield  {journal} {\bibinfo  {journal} {Nucl.
  Fusion}\ }\textbf {\bibinfo {volume} {61}},\ \bibinfo {pages} {066031}
  (\bibinfo {year} {2021})}\BibitemShut {NoStop}%
\bibitem [{\citenamefont {Chu}\ \emph {et~al.}(1992)\citenamefont {Chu},
  \citenamefont {Greene}, \citenamefont {Lao}, \citenamefont {Turnbull},\ and\
  \citenamefont {Chance}}]{chu.1992}%
  \BibitemOpen
  \bibfield  {author} {\bibinfo {author} {\bibfnamefont {M.~S.}\ \bibnamefont
  {Chu}}, \bibinfo {author} {\bibfnamefont {J.~M.}\ \bibnamefont {Greene}},
  \bibinfo {author} {\bibfnamefont {L.~L.}\ \bibnamefont {Lao}}, \bibinfo
  {author} {\bibfnamefont {A.~D.}\ \bibnamefont {Turnbull}}, \ and\ \bibinfo
  {author} {\bibfnamefont {M.~S.}\ \bibnamefont {Chance}},\ }\href {\doibase
  10.1063/1.860327} {\bibfield  {journal} {\bibinfo  {journal} {Phys. Fluids
  B}\ }\textbf {\bibinfo {volume} {4}},\ \bibinfo {pages} {3713} (\bibinfo
  {year} {1992})}\BibitemShut {NoStop}%
\bibitem [{\citenamefont {Cheng}\ \emph {et~al.}(2019)\citenamefont {Cheng},
  \citenamefont {Kramer}, \citenamefont {Podesta},\ and\ \citenamefont
  {Nazikian}}]{cheng.2019}%
  \BibitemOpen
  \bibfield  {author} {\bibinfo {author} {\bibfnamefont {C.~Z.}\ \bibnamefont
  {Cheng}}, \bibinfo {author} {\bibfnamefont {G.~J.}\ \bibnamefont {Kramer}},
  \bibinfo {author} {\bibfnamefont {M.}~\bibnamefont {Podesta}}, \ and\
  \bibinfo {author} {\bibfnamefont {R.}~\bibnamefont {Nazikian}},\ }\href
  {\doibase 10.1063/1.5108505} {\bibfield  {journal} {\bibinfo  {journal}
  {Phys. Plasmas}\ }\textbf {\bibinfo {volume} {26}},\ \bibinfo {pages}
  {082508} (\bibinfo {year} {2019})}\BibitemShut {NoStop}%
\bibitem [{\citenamefont {Kramer}\ \emph {et~al.}(2020)\citenamefont {Kramer},
  \citenamefont {Cheng}, \citenamefont {Podest{\`{a}}},\ and\ \citenamefont
  {Nazikian}}]{kramer.2020}%
  \BibitemOpen
  \bibfield  {author} {\bibinfo {author} {\bibfnamefont {G.~J.}\ \bibnamefont
  {Kramer}}, \bibinfo {author} {\bibfnamefont {C.~Z.}\ \bibnamefont {Cheng}},
  \bibinfo {author} {\bibfnamefont {M.}~\bibnamefont {Podest{\`{a}}}}, \ and\
  \bibinfo {author} {\bibfnamefont {R.}~\bibnamefont {Nazikian}},\ }\href
  {\doibase 10.1088/1361-6587/ab9153} {\bibfield  {journal} {\bibinfo
  {journal} {Plasma Phys. Control. Fusion}\ }\textbf {\bibinfo {volume} {62}},\
  \bibinfo {pages} {075012} (\bibinfo {year} {2020})}\BibitemShut {NoStop}%
\bibitem [{\citenamefont {van~der Holst}\ \emph {et~al.}(2000)\citenamefont
  {van~der Holst}, \citenamefont {Beli{\"e}n},\ and\ \citenamefont
  {Goedbloed}}]{holst.2000b}%
  \BibitemOpen
  \bibfield  {author} {\bibinfo {author} {\bibfnamefont {B.}~\bibnamefont
  {van~der Holst}}, \bibinfo {author} {\bibfnamefont {A.~J.~C.}\ \bibnamefont
  {Beli{\"e}n}}, \ and\ \bibinfo {author} {\bibfnamefont {J.~P.}\ \bibnamefont
  {Goedbloed}},\ }\href {\doibase 10.1063/1.1308084} {\bibfield  {journal}
  {\bibinfo  {journal} {Phys. Plasmas}\ }\textbf {\bibinfo {volume} {7}},\
  \bibinfo {pages} {4208} (\bibinfo {year} {2000})}\BibitemShut {NoStop}%
\bibitem [{\citenamefont {Gorelenkov}\ \emph
  {et~al.}(2007{\natexlab{a}})\citenamefont {Gorelenkov}, \citenamefont {Berk},
  \citenamefont {Fredrickson}, \citenamefont {Sharapov},\ and\ \citenamefont
  {Contributors}}]{gorelenkov.2007}%
  \BibitemOpen
  \bibfield  {author} {\bibinfo {author} {\bibfnamefont {N.}~\bibnamefont
  {Gorelenkov}}, \bibinfo {author} {\bibfnamefont {H.}~\bibnamefont {Berk}},
  \bibinfo {author} {\bibfnamefont {E.}~\bibnamefont {Fredrickson}}, \bibinfo
  {author} {\bibfnamefont {S.}~\bibnamefont {Sharapov}}, \ and\ \bibinfo
  {author} {\bibfnamefont {J.~E.}\ \bibnamefont {Contributors}},\ }\href
  {\doibase https://doi.org/10.1016/j.physleta.2007.05.113} {\bibfield
  {journal} {\bibinfo  {journal} {Phys. Lett. A}\ }\textbf {\bibinfo {volume}
  {370}},\ \bibinfo {pages} {70 } (\bibinfo {year}
  {2007}{\natexlab{a}})}\BibitemShut {NoStop}%
\bibitem [{\citenamefont {Gorelenkov}\ \emph
  {et~al.}(2007{\natexlab{b}})\citenamefont {Gorelenkov}, \citenamefont {Berk},
  \citenamefont {Crocker}, \citenamefont {Fredrickson}, \citenamefont {Kaye},
  \citenamefont {Kubota}, \citenamefont {Park}, \citenamefont {Peebles},
  \citenamefont {Sabbagh}, \citenamefont {Sharapov}, \citenamefont {Stutmat},
  \citenamefont {Tritz}, \citenamefont {Levinton},\ and\ \citenamefont
  {and}}]{gorelenkov.2007b}%
  \BibitemOpen
  \bibfield  {author} {\bibinfo {author} {\bibfnamefont {N.~N.}\ \bibnamefont
  {Gorelenkov}}, \bibinfo {author} {\bibfnamefont {H.~L.}\ \bibnamefont
  {Berk}}, \bibinfo {author} {\bibfnamefont {N.~A.}\ \bibnamefont {Crocker}},
  \bibinfo {author} {\bibfnamefont {E.~D.}\ \bibnamefont {Fredrickson}},
  \bibinfo {author} {\bibfnamefont {S.}~\bibnamefont {Kaye}}, \bibinfo {author}
  {\bibfnamefont {S.}~\bibnamefont {Kubota}}, \bibinfo {author} {\bibfnamefont
  {H.}~\bibnamefont {Park}}, \bibinfo {author} {\bibfnamefont {W.}~\bibnamefont
  {Peebles}}, \bibinfo {author} {\bibfnamefont {S.~A.}\ \bibnamefont
  {Sabbagh}}, \bibinfo {author} {\bibfnamefont {S.~E.}\ \bibnamefont
  {Sharapov}}, \bibinfo {author} {\bibfnamefont {D.}~\bibnamefont {Stutmat}},
  \bibinfo {author} {\bibfnamefont {K.}~\bibnamefont {Tritz}}, \bibinfo
  {author} {\bibfnamefont {F.~M.}\ \bibnamefont {Levinton}}, \ and\ \bibinfo
  {author} {\bibfnamefont {H.~Y.}\ \bibnamefont {and}},\ }\href {\doibase
  10.1088/0741-3335/49/12b/s34} {\bibfield  {journal} {\bibinfo  {journal}
  {Plasma Phys. Control. Fusion}\ }\textbf {\bibinfo {volume} {49}},\ \bibinfo
  {pages} {B371} (\bibinfo {year} {2007}{\natexlab{b}})}\BibitemShut {NoStop}%
\bibitem [{\citenamefont {Winsor}\ \emph {et~al.}(1968)\citenamefont {Winsor},
  \citenamefont {Johnson},\ and\ \citenamefont {Dawson}}]{winsor.1968}%
  \BibitemOpen
  \bibfield  {author} {\bibinfo {author} {\bibfnamefont {N.}~\bibnamefont
  {Winsor}}, \bibinfo {author} {\bibfnamefont {J.~L.}\ \bibnamefont {Johnson}},
  \ and\ \bibinfo {author} {\bibfnamefont {J.~M.}\ \bibnamefont {Dawson}},\
  }\href {\doibase 10.1063/1.1691835} {\bibfield  {journal} {\bibinfo
  {journal} {Phys. Fluids}\ }\textbf {\bibinfo {volume} {11}},\ \bibinfo
  {pages} {2448} (\bibinfo {year} {1968})}\BibitemShut {NoStop}%
\bibitem [{\citenamefont {Rodrigues}\ \emph {et~al.}(2021)\citenamefont
  {Rodrigues}, \citenamefont {Borba}, \citenamefont {Cella}, \citenamefont
  {Coelho}, \citenamefont {Ferreira}, \citenamefont {Figueiredo}, \citenamefont
  {Mantsinen}, \citenamefont {Nabais}, \citenamefont {Sharapov}, \citenamefont
  {Sir\'{e}n},\ and\ \citenamefont {Contributors}}]{rodrigues.2021fec}%
  \BibitemOpen
  \bibfield  {author} {\bibinfo {author} {\bibfnamefont {P.}~\bibnamefont
  {Rodrigues}}, \bibinfo {author} {\bibfnamefont {D.}~\bibnamefont {Borba}},
  \bibinfo {author} {\bibfnamefont {F.}~\bibnamefont {Cella}}, \bibinfo
  {author} {\bibfnamefont {R.}~\bibnamefont {Coelho}}, \bibinfo {author}
  {\bibfnamefont {J.}~\bibnamefont {Ferreira}}, \bibinfo {author}
  {\bibfnamefont {A.}~\bibnamefont {Figueiredo}}, \bibinfo {author}
  {\bibfnamefont {M.}~\bibnamefont {Mantsinen}}, \bibinfo {author}
  {\bibfnamefont {F.}~\bibnamefont {Nabais}}, \bibinfo {author} {\bibfnamefont
  {S.}~\bibnamefont {Sharapov}}, \bibinfo {author} {\bibfnamefont
  {P.}~\bibnamefont {Sir\'{e}n}}, \ and\ \bibinfo {author} {\bibfnamefont
  {JET}~\bibnamefont {Contributors}},\ }in\ \href
{https://nucleus.iaea.org/sites/fusionportal/Shared%20Documents/FEC%202020/fec2020-preprints/preprint1332.pdf}
{\emph {\bibinfo {booktitle} {28th IAEA Fusion Energy Conference}}}\ (\bibinfo
{address} {Nice, France, 10--15 May},\ \bibinfo {year} {2021})\BibitemShut
{NoStop}%
\bibitem [{\citenamefont {Rodrigues}\ and\ \citenamefont
  {Coroado}(2018)}]{rodrigues.2018}%
  \BibitemOpen
  \bibfield  {author} {\bibinfo {author} {\bibfnamefont {P.}~\bibnamefont
  {Rodrigues}}\ and\ \bibinfo {author} {\bibfnamefont {A.}~\bibnamefont
  {Coroado}},\ }\href {\doibase 10.1088/1741-4326/aada57} {\bibfield  {journal}
  {\bibinfo  {journal} {Nucl. Fusion}\ }\textbf {\bibinfo {volume} {58}},\
  \bibinfo {pages} {106040} (\bibinfo {year} {2018})}\BibitemShut {NoStop}%
\bibitem [{\citenamefont {Goedbloed}\ \emph {et~al.}(1993)\citenamefont
  {Goedbloed}, \citenamefont {Holties}, \citenamefont {Poedts}, \citenamefont
  {Huysmans},\ and\ \citenamefont {Kerner}}]{goedbloed.1993}%
  \BibitemOpen
  \bibfield  {author} {\bibinfo {author} {\bibfnamefont {J.~P.}\ \bibnamefont
  {Goedbloed}}, \bibinfo {author} {\bibfnamefont {H.~A.}\ \bibnamefont
  {Holties}}, \bibinfo {author} {\bibfnamefont {S.}~\bibnamefont {Poedts}},
  \bibinfo {author} {\bibfnamefont {G.~T.~A.}\ \bibnamefont {Huysmans}}, \ and\
  \bibinfo {author} {\bibfnamefont {W.}~\bibnamefont {Kerner}},\ }\href
  {\doibase 10.1088/0741-3335/35/sb/023} {\bibfield  {journal} {\bibinfo
  {journal} {Plasma Phys. Control. Fusion}\ }\textbf {\bibinfo {volume} {35}},\
  \bibinfo {pages} {B277} (\bibinfo {year} {1993})}\BibitemShut {NoStop}%
\bibitem [{\citenamefont {Fasoli}\ \emph {et~al.}(2002)\citenamefont {Fasoli},
  \citenamefont {Testa}, \citenamefont {Sharapov}, \citenamefont {Berk},
  \citenamefont {Breizman}, \citenamefont {Gondhalekar}, \citenamefont
  {Heeter}, \citenamefont {Mantsinen},\ and\ \citenamefont {contributors to~the
  EFDA-JET~Workprogramme}}]{fasoli.2002}%
  \BibitemOpen
  \bibfield  {author} {\bibinfo {author} {\bibfnamefont {A.}~\bibnamefont
  {Fasoli}}, \bibinfo {author} {\bibfnamefont {D.}~\bibnamefont {Testa}},
  \bibinfo {author} {\bibfnamefont {S.}~\bibnamefont {Sharapov}}, \bibinfo
  {author} {\bibfnamefont {H.~L.}\ \bibnamefont {Berk}}, \bibinfo {author}
  {\bibfnamefont {B.}~\bibnamefont {Breizman}}, \bibinfo {author}
  {\bibfnamefont {A.}~\bibnamefont {Gondhalekar}}, \bibinfo {author}
  {\bibfnamefont {R.~F.}\ \bibnamefont {Heeter}}, \bibinfo {author}
  {\bibfnamefont {M.}~\bibnamefont {Mantsinen}}, \ and\ \bibinfo {author}
  {\bibnamefont {contributors to~the EFDA-JET~Workprogramme}},\ }\href
  {\doibase 10.1088/0741-3335/44/12b/312} {\bibfield  {journal} {\bibinfo
  {journal} {Plasma Phys. Control. Fusion}\ }\textbf {\bibinfo {volume} {44}},\
  \bibinfo {pages} {B159} (\bibinfo {year} {2002})}\BibitemShut {NoStop}%
\bibitem [{\citenamefont {Aymar}\ \emph {et~al.}(2002)\citenamefont {Aymar},
  \citenamefont {Barabaschi},\ and\ \citenamefont {Shimomura}}]{aymar.2002}%
  \BibitemOpen
  \bibfield  {author} {\bibinfo {author} {\bibfnamefont {R.}~\bibnamefont
  {Aymar}}, \bibinfo {author} {\bibfnamefont {P.}~\bibnamefont {Barabaschi}}, \
  and\ \bibinfo {author} {\bibfnamefont {Y.}~\bibnamefont {Shimomura}},\ }\href
  {\doibase 10.1088/0741-3335/44/5/304} {\bibfield  {journal} {\bibinfo
  {journal} {Plasma Phys. Control. Fusion}\ }\textbf {\bibinfo {volume} {44}},\
  \bibinfo {pages} {519} (\bibinfo {year} {2002})}\BibitemShut {NoStop}%
\bibitem [{\citenamefont {Hameiri}(1981)}]{hameiri.1981}%
  \BibitemOpen
  \bibfield  {author} {\bibinfo {author} {\bibfnamefont {E.}~\bibnamefont
  {Hameiri}},\ }\href {\doibase 10.1063/1.863410} {\bibfield  {journal}
  {\bibinfo  {journal} {Phys. Fluids}\ }\textbf {\bibinfo {volume} {24}},\
  \bibinfo {pages} {562} (\bibinfo {year} {1981})}\BibitemShut {NoStop}%
\bibitem [{\citenamefont {Hameiri}(1985)}]{hameiri.1985}%
  \BibitemOpen
  \bibfield  {author} {\bibinfo {author} {\bibfnamefont {E.}~\bibnamefont
  {Hameiri}},\ }\href {\doibase 10.1002/cpa.3160380104} {\bibfield  {journal}
  {\bibinfo  {journal} {Comm. Pure Appl. Math.}\ }\textbf {\bibinfo {volume}
  {38}},\ \bibinfo {pages} {43} (\bibinfo {year} {1985})}\BibitemShut {NoStop}%
\bibitem [{\citenamefont {Dumont}\ \emph {et~al.}(2018)\citenamefont {Dumont},
  \citenamefont {Mailloux}, \citenamefont {Aslanyan}, \citenamefont {Baruzzo},
  \citenamefont {Challis}, \citenamefont {Coffey}, \citenamefont {Czarnecka},
  \citenamefont {Delabie}, \citenamefont {Eriksson}, \citenamefont {Faustin},
  \citenamefont {Ferreira}, \citenamefont {Fitzgerald}, \citenamefont {Garcia},
  \citenamefont {Giacomelli}, \citenamefont {Giroud}, \citenamefont {Hawkes},
  \citenamefont {Jacquet}, \citenamefont {Joffrin}, \citenamefont {Johnson},
  \citenamefont {Keeling}, \citenamefont {King}, \citenamefont {Kiptily},
  \citenamefont {Lomanowski}, \citenamefont {Lerche}, \citenamefont
  {Mantsinen}, \citenamefont {Meneses}, \citenamefont {Menmuir}, \citenamefont
  {McClements}, \citenamefont {Moradi}, \citenamefont {Nabais}, \citenamefont
  {Nocente}, \citenamefont {Patel}, \citenamefont {Patten}, \citenamefont
  {Puglia}, \citenamefont {Scannell}, \citenamefont {Sharapov}, \citenamefont
  {Solano}, \citenamefont {Tsalas}, \citenamefont {Vallejos},\ and\
  \citenamefont {and}}]{dumont.2018}%
  \BibitemOpen
  \bibfield  {author} {\bibinfo {author} {\bibfnamefont {R.~J.}\ \bibnamefont
  {Dumont}}, \bibinfo {author} {\bibfnamefont {J.}~\bibnamefont {Mailloux}},
  \bibinfo {author} {\bibfnamefont {V.}~\bibnamefont {Aslanyan}}, \bibinfo
  {author} {\bibfnamefont {M.}~\bibnamefont {Baruzzo}}, \bibinfo {author}
  {\bibfnamefont {C.}~\bibnamefont {Challis}}, \bibinfo {author} {\bibfnamefont
  {I.}~\bibnamefont {Coffey}}, \bibinfo {author} {\bibfnamefont
  {A.}~\bibnamefont {Czarnecka}}, \bibinfo {author} {\bibfnamefont
  {E.}~\bibnamefont {Delabie}}, \bibinfo {author} {\bibfnamefont
  {J.}~\bibnamefont {Eriksson}}, \bibinfo {author} {\bibfnamefont
  {J.}~\bibnamefont {Faustin}}, \bibinfo {author} {\bibfnamefont
  {J.}~\bibnamefont {Ferreira}}, \bibinfo {author} {\bibfnamefont
  {M.}~\bibnamefont {Fitzgerald}}, \bibinfo {author} {\bibfnamefont
  {J.}~\bibnamefont {Garcia}}, \bibinfo {author} {\bibfnamefont
  {L.}~\bibnamefont {Giacomelli}}, \bibinfo {author} {\bibfnamefont
  {C.}~\bibnamefont {Giroud}}, \bibinfo {author} {\bibfnamefont
  {N.}~\bibnamefont {Hawkes}}, \bibinfo {author} {\bibfnamefont
  {P.}~\bibnamefont {Jacquet}}, \bibinfo {author} {\bibfnamefont
  {E.}~\bibnamefont {Joffrin}}, \bibinfo {author} {\bibfnamefont
  {T.}~\bibnamefont {Johnson}}, \bibinfo {author} {\bibfnamefont
  {D.}~\bibnamefont {Keeling}}, \bibinfo {author} {\bibfnamefont
  {D.}~\bibnamefont {King}}, \bibinfo {author} {\bibfnamefont {V.}~\bibnamefont
  {Kiptily}}, \bibinfo {author} {\bibfnamefont {B.}~\bibnamefont {Lomanowski}},
  \bibinfo {author} {\bibfnamefont {E.}~\bibnamefont {Lerche}}, \bibinfo
  {author} {\bibfnamefont {M.}~\bibnamefont {Mantsinen}}, \bibinfo {author}
  {\bibfnamefont {L.}~\bibnamefont {Meneses}}, \bibinfo {author} {\bibfnamefont
  {S.}~\bibnamefont {Menmuir}}, \bibinfo {author} {\bibfnamefont
  {K.}~\bibnamefont {McClements}}, \bibinfo {author} {\bibfnamefont
  {S.}~\bibnamefont {Moradi}}, \bibinfo {author} {\bibfnamefont
  {F.}~\bibnamefont {Nabais}}, \bibinfo {author} {\bibfnamefont
  {M.}~\bibnamefont {Nocente}}, \bibinfo {author} {\bibfnamefont
  {A.}~\bibnamefont {Patel}}, \bibinfo {author} {\bibfnamefont
  {H.}~\bibnamefont {Patten}}, \bibinfo {author} {\bibfnamefont
  {P.}~\bibnamefont {Puglia}}, \bibinfo {author} {\bibfnamefont
  {R.}~\bibnamefont {Scannell}}, \bibinfo {author} {\bibfnamefont
  {S.}~\bibnamefont {Sharapov}}, \bibinfo {author} {\bibfnamefont {E.~R.}\
  \bibnamefont {Solano}}, \bibinfo {author} {\bibfnamefont {M.}~\bibnamefont
  {Tsalas}}, \bibinfo {author} {\bibfnamefont {P.}~\bibnamefont {Vallejos}}, \
  and\ \bibinfo {author} {\bibfnamefont {H.~W.}\ \bibnamefont {and}},\ }\href
  {\doibase 10.1088/1741-4326/aab1bb} {\bibfield  {journal} {\bibinfo
  {journal} {Nucl. Fusion}\ }\textbf {\bibinfo {volume} {58}},\ \bibinfo
  {pages} {082005} (\bibinfo {year} {2018})}\BibitemShut {NoStop}%
\bibitem [{\citenamefont {Huysmans}\ \emph {et~al.}(1991)\citenamefont
  {Huysmans}, \citenamefont {Goedbloed},\ and\ \citenamefont
  {Kerner}}]{huysmans.1991}%
  \BibitemOpen
  \bibfield  {author} {\bibinfo {author} {\bibfnamefont {G.}~\bibnamefont
  {Huysmans}}, \bibinfo {author} {\bibfnamefont {J.}~\bibnamefont {Goedbloed}},
  \ and\ \bibinfo {author} {\bibfnamefont {W.}~\bibnamefont {Kerner}},\ }\href
  {\doibase 10.1142/S0129183191000512} {\bibfield  {journal} {\bibinfo
  {journal} {Int. J. Mod. Phys. C}\ }\textbf {\bibinfo {volume} {2}},\ \bibinfo
  {pages} {371} (\bibinfo {year} {1991})}\BibitemShut {NoStop}%
\bibitem [{\citenamefont {Poedts}\ and\ \citenamefont
  {Schwartz}(1993)}]{poedts.1993}%
  \BibitemOpen
  \bibfield  {author} {\bibinfo {author} {\bibfnamefont {S.}~\bibnamefont
  {Poedts}}\ and\ \bibinfo {author} {\bibfnamefont {E.}~\bibnamefont
  {Schwartz}},\ }\href {\doibase https://doi.org/10.1006/jcph.1993.1061}
  {\bibfield  {journal} {\bibinfo  {journal} {J. Comput. Phys.}\ }\textbf
  {\bibinfo {volume} {105}},\ \bibinfo {pages} {165 } (\bibinfo {year}
  {1993})}\BibitemShut {NoStop}%
\bibitem [{\citenamefont {Kerner}\ \emph {et~al.}(1998)\citenamefont {Kerner},
  \citenamefont {Goedbloed}, \citenamefont {Huysmans}, \citenamefont {Poedts},\
  and\ \citenamefont {Schwarz}}]{kerner.1998}%
  \BibitemOpen
  \bibfield  {author} {\bibinfo {author} {\bibfnamefont {W.}~\bibnamefont
  {Kerner}}, \bibinfo {author} {\bibfnamefont {J.}~\bibnamefont {Goedbloed}},
  \bibinfo {author} {\bibfnamefont {G.}~\bibnamefont {Huysmans}}, \bibinfo
  {author} {\bibfnamefont {S.}~\bibnamefont {Poedts}}, \ and\ \bibinfo {author}
  {\bibfnamefont {E.}~\bibnamefont {Schwarz}},\ }\href {\doibase
  10.1006/jcph.1998.5910} {\bibfield  {journal} {\bibinfo  {journal} {J.
  Comput. Phys.}\ }\textbf {\bibinfo {volume} {142}},\ \bibinfo {pages} {271 }
  (\bibinfo {year} {1998})}\BibitemShut {NoStop}%
\bibitem [{\citenamefont {Porcelli}\ \emph {et~al.}(1994)\citenamefont
  {Porcelli}, \citenamefont {Stankiewicz}, \citenamefont {Kerner},\ and\
  \citenamefont {Berk}}]{porcelli.1994}%
  \BibitemOpen
  \bibfield  {author} {\bibinfo {author} {\bibfnamefont {F.}~\bibnamefont
  {Porcelli}}, \bibinfo {author} {\bibfnamefont {R.}~\bibnamefont
  {Stankiewicz}}, \bibinfo {author} {\bibfnamefont {W.}~\bibnamefont {Kerner}},
  \ and\ \bibinfo {author} {\bibfnamefont {H.~L.}\ \bibnamefont {Berk}},\
  }\href {\doibase 10.1063/1.870792} {\bibfield  {journal} {\bibinfo  {journal}
  {Phys. Plasmas}\ }\textbf {\bibinfo {volume} {1}},\ \bibinfo {pages} {470}
  (\bibinfo {year} {1994})}\BibitemShut {NoStop}%
\bibitem [{\citenamefont {Borba}\ and\ \citenamefont
  {Kerner}(1999)}]{borba.1999}%
  \BibitemOpen
  \bibfield  {author} {\bibinfo {author} {\bibfnamefont {D.}~\bibnamefont
  {Borba}}\ and\ \bibinfo {author} {\bibfnamefont {W.}~\bibnamefont {Kerner}},\
  }\href {\doibase 10.1006/jcph.1999.6264} {\bibfield  {journal} {\bibinfo
  {journal} {J. Comput. Phys.}\ }\textbf {\bibinfo {volume} {153}},\ \bibinfo
  {pages} {101 } (\bibinfo {year} {1999})}\BibitemShut {NoStop}%
\bibitem [{\citenamefont {Nabais}\ \emph {et~al.}(2015)\citenamefont {Nabais},
  \citenamefont {Borba}, \citenamefont {Coelho}, \citenamefont {Figueiredo},
  \citenamefont {Ferreira}, \citenamefont {Loureiro},\ and\ \citenamefont
  {Rodrigues}}]{nabais.2015}%
  \BibitemOpen
  \bibfield  {author} {\bibinfo {author} {\bibfnamefont {F.}~\bibnamefont
  {Nabais}}, \bibinfo {author} {\bibfnamefont {D.}~\bibnamefont {Borba}},
  \bibinfo {author} {\bibfnamefont {R.}~\bibnamefont {Coelho}}, \bibinfo
  {author} {\bibfnamefont {A.}~\bibnamefont {Figueiredo}}, \bibinfo {author}
  {\bibfnamefont {J.}~\bibnamefont {Ferreira}}, \bibinfo {author}
  {\bibfnamefont {N.}~\bibnamefont {Loureiro}}, \ and\ \bibinfo {author}
  {\bibfnamefont {P.}~\bibnamefont {Rodrigues}},\ }\href {\doibase
  10.1088/1009-0630/17/2/01} {\bibfield  {journal} {\bibinfo  {journal} {Plasma
  Sci. Technol.}\ }\textbf {\bibinfo {volume} {17}},\ \bibinfo {pages} {89}
  (\bibinfo {year} {2015})}\BibitemShut {NoStop}%
\bibitem [{\citenamefont {Gorelenkov}\ and\ \citenamefont
  {Sharapov}(1992)}]{gorelenkov.1992}%
  \BibitemOpen
  \bibfield  {author} {\bibinfo {author} {\bibfnamefont {N.~N.}\ \bibnamefont
  {Gorelenkov}}\ and\ \bibinfo {author} {\bibfnamefont {S.~E.}\ \bibnamefont
  {Sharapov}},\ }\href {\doibase 10.1088/0031-8949/45/2/016} {\bibfield
  {journal} {\bibinfo  {journal} {Phys. Scr.}\ }\textbf {\bibinfo {volume}
  {45}},\ \bibinfo {pages} {163} (\bibinfo {year} {1992})}\BibitemShut
  {NoStop}%
\bibitem [{\citenamefont {Mett}\ and\ \citenamefont
  {Mahajan}(1992)}]{mett.1992}%
  \BibitemOpen
  \bibfield  {author} {\bibinfo {author} {\bibfnamefont {R.~R.}\ \bibnamefont
  {Mett}}\ and\ \bibinfo {author} {\bibfnamefont {S.~M.}\ \bibnamefont
  {Mahajan}},\ }\href {\doibase 10.1063/1.860459} {\bibfield  {journal}
  {\bibinfo  {journal} {Phys. Fluids B}\ }\textbf {\bibinfo {volume} {4}},\
  \bibinfo {pages} {2885} (\bibinfo {year} {1992})}\BibitemShut {NoStop}%
\bibitem [{\citenamefont {Candy}\ and\ \citenamefont
  {Rosenbluth}(1994)}]{candy.1994}%
  \BibitemOpen
  \bibfield  {author} {\bibinfo {author} {\bibfnamefont {J.}~\bibnamefont
  {Candy}}\ and\ \bibinfo {author} {\bibfnamefont {M.~N.}\ \bibnamefont
  {Rosenbluth}},\ }\href {\doibase 10.1063/1.870838} {\bibfield  {journal}
  {\bibinfo  {journal} {Phys. Plasmas}\ }\textbf {\bibinfo {volume} {1}},\
  \bibinfo {pages} {356} (\bibinfo {year} {1994})}\BibitemShut {NoStop}%
\bibitem [{\citenamefont {Rosenbluth}\ \emph {et~al.}(1992)\citenamefont
  {Rosenbluth}, \citenamefont {Berk}, \citenamefont {Van~Dam},\ and\
  \citenamefont {Lindberg}}]{rosenbluth.1992}%
  \BibitemOpen
  \bibfield  {author} {\bibinfo {author} {\bibfnamefont {M.~N.}\ \bibnamefont
  {Rosenbluth}}, \bibinfo {author} {\bibfnamefont {H.~L.}\ \bibnamefont
  {Berk}}, \bibinfo {author} {\bibfnamefont {J.~W.}\ \bibnamefont {Van~Dam}}, \
  and\ \bibinfo {author} {\bibfnamefont {D.~M.}\ \bibnamefont {Lindberg}},\
  }\href {\doibase 10.1103/PhysRevLett.68.596} {\bibfield  {journal} {\bibinfo
  {journal} {Phys. Rev. Lett.}\ }\textbf {\bibinfo {volume} {68}},\ \bibinfo
  {pages} {596} (\bibinfo {year} {1992})}\BibitemShut {NoStop}%
\bibitem [{\citenamefont {Zonca}\ and\ \citenamefont
  {Chen}(1992)}]{zonca.1992}%
  \BibitemOpen
  \bibfield  {author} {\bibinfo {author} {\bibfnamefont {F.}~\bibnamefont
  {Zonca}}\ and\ \bibinfo {author} {\bibfnamefont {L.}~\bibnamefont {Chen}},\
  }\href {\doibase 10.1103/PhysRevLett.68.592} {\bibfield  {journal} {\bibinfo
  {journal} {Phys. Rev. Lett.}\ }\textbf {\bibinfo {volume} {68}},\ \bibinfo
  {pages} {592} (\bibinfo {year} {1992})}\BibitemShut {NoStop}%
\bibitem [{\citenamefont {Poedts}\ and\ \citenamefont
  {Kerner}(1991)}]{poedts.1991}%
  \BibitemOpen
  \bibfield  {author} {\bibinfo {author} {\bibfnamefont {S.}~\bibnamefont
  {Poedts}}\ and\ \bibinfo {author} {\bibfnamefont {W.}~\bibnamefont
  {Kerner}},\ }\href {\doibase 10.1103/PhysRevLett.66.2871} {\bibfield
  {journal} {\bibinfo  {journal} {Phys. Rev. Lett.}\ }\textbf {\bibinfo
  {volume} {66}},\ \bibinfo {pages} {2871} (\bibinfo {year}
  {1991})}\BibitemShut {NoStop}%
\bibitem [{\citenamefont {Poedts}\ \emph {et~al.}(1992)\citenamefont {Poedts},
  \citenamefont {Kerner}, \citenamefont {Goedbloed}, \citenamefont {Keegan},
  \citenamefont {Huysmans},\ and\ \citenamefont {Schwarz}}]{poedts.1992}%
  \BibitemOpen
  \bibfield  {author} {\bibinfo {author} {\bibfnamefont {S.}~\bibnamefont
  {Poedts}}, \bibinfo {author} {\bibfnamefont {W.}~\bibnamefont {Kerner}},
  \bibinfo {author} {\bibfnamefont {J.~P.}\ \bibnamefont {Goedbloed}}, \bibinfo
  {author} {\bibfnamefont {B.}~\bibnamefont {Keegan}}, \bibinfo {author}
  {\bibfnamefont {G.~T.~A.}\ \bibnamefont {Huysmans}}, \ and\ \bibinfo {author}
  {\bibfnamefont {E.}~\bibnamefont {Schwarz}},\ }\href {\doibase
  10.1088/0741-3335/34/8/003} {\bibfield  {journal} {\bibinfo  {journal}
  {Plasma Phys. Control. Fusion}\ }\textbf {\bibinfo {volume} {34}},\ \bibinfo
  {pages} {1397} (\bibinfo {year} {1992})}\BibitemShut {NoStop}%
\bibitem [{\citenamefont {Todo}\ and\ \citenamefont
  {Bierwage}(2014)}]{todo.2014}%
  \BibitemOpen
  \bibfield  {author} {\bibinfo {author} {\bibfnamefont {Y.}~\bibnamefont
  {Todo}}\ and\ \bibinfo {author} {\bibfnamefont {A.}~\bibnamefont
  {Bierwage}},\ }\href {\doibase 10.1585/pfr.9.3403068} {\bibfield  {journal}
  {\bibinfo  {journal} {Plasma Fusion Res.}\ }\textbf {\bibinfo {volume} {9}},\
  \bibinfo {pages} {3403068} (\bibinfo {year} {2014})}\BibitemShut {NoStop}%
\bibitem [{\citenamefont {Bierwage}\ \emph {et~al.}(2015)\citenamefont
  {Bierwage}, \citenamefont {Aiba},\ and\ \citenamefont
  {Shinohara}}]{bierwage.2015}%
  \BibitemOpen
  \bibfield  {author} {\bibinfo {author} {\bibfnamefont {A.}~\bibnamefont
  {Bierwage}}, \bibinfo {author} {\bibfnamefont {N.}~\bibnamefont {Aiba}}, \
  and\ \bibinfo {author} {\bibfnamefont {K.}~\bibnamefont {Shinohara}},\ }\href
  {\doibase 10.1103/PhysRevLett.114.015002} {\bibfield  {journal} {\bibinfo
  {journal} {Phys. Rev. Lett.}\ }\textbf {\bibinfo {volume} {114}},\ \bibinfo
  {pages} {015002} (\bibinfo {year} {2015})}\BibitemShut {NoStop}%
\bibitem [{\citenamefont {Pinches}\ \emph {et~al.}(2015)\citenamefont
  {Pinches}, \citenamefont {Chapman}, \citenamefont {Lauber}, \citenamefont
  {Oliver}, \citenamefont {Sharapov}, \citenamefont {Shinohara},\ and\
  \citenamefont {Tani}}]{pinches.2015}%
  \BibitemOpen
  \bibfield  {author} {\bibinfo {author} {\bibfnamefont {S.~D.}\ \bibnamefont
  {Pinches}}, \bibinfo {author} {\bibfnamefont {I.~T.}\ \bibnamefont
  {Chapman}}, \bibinfo {author} {\bibfnamefont {P.~W.}\ \bibnamefont {Lauber}},
  \bibinfo {author} {\bibfnamefont {H.~J.~C.}\ \bibnamefont {Oliver}}, \bibinfo
  {author} {\bibfnamefont {S.~E.}\ \bibnamefont {Sharapov}}, \bibinfo {author}
  {\bibfnamefont {K.}~\bibnamefont {Shinohara}}, \ and\ \bibinfo {author}
  {\bibfnamefont {K.}~\bibnamefont {Tani}},\ }\href {\doibase
  10.1063/1.4908551} {\bibfield  {journal} {\bibinfo  {journal} {Phys.
  Plasmas}\ }\textbf {\bibinfo {volume} {22}},\ \bibinfo {pages} {021807}
  (\bibinfo {year} {2015})}\BibitemShut {NoStop}%
\bibitem [{\citenamefont {Lauber}(2015)}]{lauber.2015}%
  \BibitemOpen
  \bibfield  {author} {\bibinfo {author} {\bibfnamefont {P.}~\bibnamefont
  {Lauber}},\ }\href {\doibase 10.1088/0741-3335/57/5/054011} {\bibfield
  {journal} {\bibinfo  {journal} {Plasma Phys. Control. Fusion}\ }\textbf
  {\bibinfo {volume} {57}},\ \bibinfo {pages} {054011} (\bibinfo {year}
  {2015})}\BibitemShut {NoStop}%
\bibitem [{\citenamefont {Rodrigues}\ \emph {et~al.}(2015)\citenamefont
  {Rodrigues}, \citenamefont {Figueiredo}, \citenamefont {Ferreira},
  \citenamefont {Coelho}, \citenamefont {Nabais}, \citenamefont {Borba},
  \citenamefont {Loureiro}, \citenamefont {Oliver},\ and\ \citenamefont
  {Sharapov}}]{rodrigues.2015}%
  \BibitemOpen
  \bibfield  {author} {\bibinfo {author} {\bibfnamefont {P.}~\bibnamefont
  {Rodrigues}}, \bibinfo {author} {\bibfnamefont {A.}~\bibnamefont
  {Figueiredo}}, \bibinfo {author} {\bibfnamefont {J.}~\bibnamefont
  {Ferreira}}, \bibinfo {author} {\bibfnamefont {R.}~\bibnamefont {Coelho}},
  \bibinfo {author} {\bibfnamefont {F.}~\bibnamefont {Nabais}}, \bibinfo
  {author} {\bibfnamefont {D.}~\bibnamefont {Borba}}, \bibinfo {author}
  {\bibfnamefont {N.}~\bibnamefont {Loureiro}}, \bibinfo {author}
  {\bibfnamefont {H.}~\bibnamefont {Oliver}}, \ and\ \bibinfo {author}
  {\bibfnamefont {S.}~\bibnamefont {Sharapov}},\ }\href {\doibase
  10.1088/0029-5515/55/8/083003} {\bibfield  {journal} {\bibinfo  {journal}
  {Nucl. Fusion}\ }\textbf {\bibinfo {volume} {55}},\ \bibinfo {pages} {083003}
  (\bibinfo {year} {2015})}\BibitemShut {NoStop}%
\bibitem [{\citenamefont {Figueiredo}\ \emph {et~al.}(2016)\citenamefont
  {Figueiredo}, \citenamefont {Rodrigues}, \citenamefont {Borba}, \citenamefont
  {Coelho}, \citenamefont {Fazendeiro}, \citenamefont {Ferreira}, \citenamefont
  {Loureiro}, \citenamefont {Nabais}, \citenamefont {Pinches}, \citenamefont
  {Polevoi},\ and\ \citenamefont {Sharapov}}]{figueiredo.2016}%
  \BibitemOpen
  \bibfield  {author} {\bibinfo {author} {\bibfnamefont {A.}~\bibnamefont
  {Figueiredo}}, \bibinfo {author} {\bibfnamefont {P.}~\bibnamefont
  {Rodrigues}}, \bibinfo {author} {\bibfnamefont {D.}~\bibnamefont {Borba}},
  \bibinfo {author} {\bibfnamefont {R.}~\bibnamefont {Coelho}}, \bibinfo
  {author} {\bibfnamefont {L.}~\bibnamefont {Fazendeiro}}, \bibinfo {author}
  {\bibfnamefont {J.}~\bibnamefont {Ferreira}}, \bibinfo {author}
  {\bibfnamefont {N.}~\bibnamefont {Loureiro}}, \bibinfo {author}
  {\bibfnamefont {F.}~\bibnamefont {Nabais}}, \bibinfo {author} {\bibfnamefont
  {S.}~\bibnamefont {Pinches}}, \bibinfo {author} {\bibfnamefont
  {A.}~\bibnamefont {Polevoi}}, \ and\ \bibinfo {author} {\bibfnamefont
  {S.}~\bibnamefont {Sharapov}},\ }\href {\doibase
  10.1088/0029-5515/56/7/076007} {\bibfield  {journal} {\bibinfo  {journal}
  {Nucl. Fusion}\ }\textbf {\bibinfo {volume} {56}},\ \bibinfo {pages} {076007}
  (\bibinfo {year} {2016})}\BibitemShut {NoStop}%
\bibitem [{\citenamefont {Fitzgerald}\ \emph {et~al.}(2016)\citenamefont
  {Fitzgerald}, \citenamefont {Sharapov}, \citenamefont {Rodrigues},\ and\
  \citenamefont {Borba}}]{fitzgerald.2016}%
  \BibitemOpen
  \bibfield  {author} {\bibinfo {author} {\bibfnamefont {M.}~\bibnamefont
  {Fitzgerald}}, \bibinfo {author} {\bibfnamefont {S.}~\bibnamefont
  {Sharapov}}, \bibinfo {author} {\bibfnamefont {P.}~\bibnamefont {Rodrigues}},
  \ and\ \bibinfo {author} {\bibfnamefont {D.}~\bibnamefont {Borba}},\ }\href
  {\doibase 10.1088/0029-5515/56/11/112010} {\bibfield  {journal} {\bibinfo
  {journal} {Nucl. Fusion}\ }\textbf {\bibinfo {volume} {56}},\ \bibinfo
  {pages} {112010} (\bibinfo {year} {2016})}\BibitemShut {NoStop}%
\bibitem [{\citenamefont {Schneller}\ \emph {et~al.}(2016)\citenamefont
  {Schneller}, \citenamefont {Lauber},\ and\ \citenamefont
  {Briguglio}}]{schneller.2016}%
  \BibitemOpen
  \bibfield  {author} {\bibinfo {author} {\bibfnamefont {M.}~\bibnamefont
  {Schneller}}, \bibinfo {author} {\bibfnamefont {P.}~\bibnamefont {Lauber}}, \
  and\ \bibinfo {author} {\bibfnamefont {S.}~\bibnamefont {Briguglio}},\ }\href
  {\doibase 10.1088/0741-3335/58/1/014019} {\bibfield  {journal} {\bibinfo
  {journal} {Plasma Phys. Control. Fusion}\ }\textbf {\bibinfo {volume} {58}},\
  \bibinfo {pages} {014019} (\bibinfo {year} {2016})}\BibitemShut {NoStop}%
\end{thebibliography}
\end{document}